\documentclass[%
 reprint,
superscriptaddress,
 amsmath,amssymb,
 aps,
]{revtex4-1}

\usepackage{graphicx}
\usepackage{dcolumn}
\usepackage{bm}

\usepackage{textcomp}
\usepackage[centerlast]{caption}
\usepackage{subcaption}
\usepackage{amssymb}
\usepackage{amsmath}

\begin{document}


\title{Autler-Townes splitting and acoustically induced transparency based on Love waves interacting with pillared meta-surface}

\author{Yuxin Liu}
\email{yuxin.liu@phd.ec-lille.fr}
\affiliation{Univ. Lille, CNRS, Centrale Lille, ISEN, Univ. Valenciennes, UMR 8520 - IEMN \& LIA LICS/LEMAC, F-59000 Lille, France}
\author{Abdelkrim Talbi}%

\affiliation{Univ. Lille, CNRS, Centrale Lille, ISEN, Univ. Valenciennes, UMR 8520 - IEMN \& LIA LICS/LEMAC, F-59000 Lille, France}

\author{El Houssaine El Boudouti}%
\affiliation{LPMR, Department of Physics, Faculty of sciences, University Mohammed I, 60000 Oujda, Morocco}

\author{Olivier Bou Matar}

\author{Philippe Pernod}
\author{Bahram Djafari-Rouhani}
\affiliation{Univ. Lille, CNRS, Centrale Lille, ISEN, Univ. Valenciennes, UMR 8520 - IEMN \& LIA LICS/LEMAC, F-59000 Lille, France}

\date{\today}

\begin{abstract}
Autler-Townes Splitting (ATS) and Electromagnetically Induced Transparency (EIT) are similar phenomena but distinct in nature. They have been widely discussed and distinguished by employing the Akaike information criterion (AIC). However, such work is lacking in acoustic system. 
In this work, the interaction of Love waves with two-line pillared meta-surface is numerically investigated by Finite Element Method. 
Acoustic analogue of ATS, Fabry-Perot resonance and cavity modes are first demonstrated in two lines of identical pillars by varying the distance between the pillar lines. 
By detuning the radius of one line of pillars, Fabry-Perot resonance along with two different pillar resonances give rise to the acoustic analogue of EIT (AIT) when the distance between the pillar lines is a multiple of half wavelength.
%
ATS and AIT formula models are used to fit the transmission spectra, showing good agreements with numerical results. The quality of the fit models is quantitatively evaluated by resorting to the AIC.  We show that theoretical and analytical discrimination between ATS and AIT are methodologically complementary.
These results should have important consequences for potential acoustic applications such as wave control, designing of meta-materials and bio-sensors.
\begin{description}
\item[PACS numbers]
68.60.Bs
\end{description}
\end{abstract}

\pacs{Valid PACS appear here}
\maketitle


\section{Introduction}
Electromagnetically induced transparency (EIT) is a well-known physical effect in atomic systems that arise because of quantum destructive interferences between two excitation pathways to an upper atomic level \cite{fleischhauer_electromagnetically_2005}. Steep dispersion and low absorption take place in a sharp transparency window, which makes it very attractive for a plenty of potential applications in slowing light, enhancing optical nonlinearity and data storage \cite{hau_light_1999,alotaibi_enhanced_2016,heinze_stopped_2013}. Autler–Townes splitting \cite{autler_stark_1955} (ATS) which is the field-induced splitting of the optical response, is not associated with interference effects and has been described as an incoherent sum of two Lorentzians \cite{abisalloum_electromagnetically_2010}.  EIT and ATS may phenomenologically look similar, but they are different in nature, one being a quantum interference and the other a linear alternating current (AC) Stark effect.
%
EIT and ATS were first observed in quantum/atom systems\cite{boller_observation_1991,autler_stark_1955}. 
In recent years, classic analogues of EIT and ATS have attracted increasing interest in platforms such as photonics \cite{peng_what_2014,liu_electromagnetically_2017,wei_objectively_2017}, optomechanics \cite{dong_optomechanical_2012,safavi_electromagnetically_2011,weis_optomechanically_2010}, plasmonics \cite{zhang_plasmon_2008,liu_plasmonic_2009,Han_plasmon_2011} and metamaterials \cite{liu_planar_2010,gu_active_2012,papasimakis_metamaterial_2008}. 
Many discussions have been devoted to their easily confused absorption or transmission spectra.\cite{anisimov_objectively_2011,peng_what_2014,sun_electromagnetically_2014,liu_experimental_2016} Besides their differences in the physical mechanisms, the Akaike Information Criterion (AIC) has been proposed to quantitatively discern EIT from ATS,\cite{anisimov_objectively_2011} and the transition from ATS to EIT is thereby carried out.\cite{giner_experimental_2013} 
A crossover from EIT to ATS has been shown to exist in hot molecules\cite{zhu_crossover_2013}, and in open ladder systems\cite{tan_crossover_2014}.
In acoustic, the analogue of EIT, also referred to as AIT, has been investigated in different structures\cite{radeonychev_acoustically_2006,liu_acoustic_2010,santillan_acoustic_2011,amin_acoustically_2015,quotane_trapped_2018}, but the analogue of ATS and its comparison with AIT was only recently reported\cite{jin_acoustic_2018}. Additionally, the distinction and transition between acoustic analogue of ATS and AIT has not been quantitatively investigated before. 

In the last two decades, phononic crystals have received increasing attention. \cite{martinez-sala_sound_1995,liu_locally_2000,khelif_trapping_2003,pennec_two-dimensional_2010,zhu_holey-structured_2011,yang_extreme_2014,xu_implementation_2018} They exhibit Bragg and/or hybridization band gaps to achieve control of elastic waves propagation, and apply to various aspects such as ratio frequency (RF) communications \cite{khelif_trapping_2003,wu_waveguiding_2009,liang_acoustic_2010}, acoustic isolators \cite{meseguer_rayleigh-wave_1999,olsson_iii_microfabricated_2009,binci_planar_2016}, sensors \cite{talbi_zno/quartz_2006,ke_sub-wavelength_2011,salman_low-concentration_2015,xu_implementation_2018}, thermoelectric materials \cite{yu_reduction_2010,yang_extreme_2014,zen_engineering_2014} and meta-materials \cite{zhang_negative_2004,profunser_dynamic_2009,zhu_holey-structured_2011,narayana_heat_2012,li_acoustic_2015,guo_acoustic_2017}. Phononic pillared meta-surface is a recently proposed structure stemming from the pillared phononic crystals. It consists of a single or a line of pillars on top of a slab, with which one can thoroughly investigate the pillar resonant properties. The pillar size, periodicity and the slab thickness are all sub-wavelength-scaled\cite{li_experimetal_2011,yu_flat_2014}. Several studies have been devoted to the interaction of pillared meta-surface with Rayleigh waves \cite{oudich_rayleigh_2018} and Lamb waves \cite{jin_tunable_2017,jin_acoustic_2017}, but no work has been done on Love waves, which are shear horizontal (SH) polarized surface acoustic waves (SAW). 
Love waves propagate in a thin guiding layer deposited on the surface of a semi-infinite substrate, and is therefore considered a compromise between Rayleigh waves and Lamb waves: Compared with Rayleigh waves, Love waves are well confined to the surface because of the thin guiding layer in which they propagate. In contrast to Lamb waves, Love waves exhibit device toughness since the guiding layer is deposited on the semi-infinite substrate. Love-wave-based pillared meta-surface is therefore suitable for use in wave control, sensors and design of metamaterials. Moreover, Love waves can excite pillar torsional mode \cite{liu_interaction_2019} which is impossible to be excited by Rayleigh waves and Lamb waves (S0 and A0) \cite{jin_tunable_2017}. In addition to the above qualities, their compatibility with liquid environment \cite{lucklum_phononic_2009,ke_sub-wavelength_2011} makes Love wave an ideal candidate for bio-sensor applications.

In this work, the interaction of Love waves with a two-line Ni pillar based meta-surface is investigated on a SiO$_{2}$/ST quartz structure. Firstly, torsional mode in one line of cylindrical Ni pillars is demonstrated to be well excited by Love waves and gives rise to a sharp transmission dip due to a destructive interference. Secondly, acoustic analogue of ATS and cavity modes for Love waves are first demonstrated in two lines of identical pillars by varying the distance between the pillar lines. ATS appears when the distance is smaller than the half wavelength and a strong coupling is aroused between the pillar lines. 
Fabry-Perot resonance exists at the positions where the distance between the pillar lines is a multiple of half wavelength. 
The proximity of Fabry-Perot resonance with pillar torsional mode gives rise to the cavity modes with transmission enhancement at the edges of the dip induced by the pillars. 
Thirdly, the radius of one line of pillar is modified to detune the pillar resonant frequency. In the pillar coupling region, the coupling effect decreases with the increase of radius mismatch. When the distance between the pillar lines is a multiple of half wavelength, Fabry-Perot resonance along with the two different pillar resonances give rise to the AIT resonance.
Then, ATS and AIT phenomena are first fitted respectively with corresponding formula models, showing good agreements. The Akaike Information Criterion (AIC) is then used to quantitatively evaluate the quality of the fit models, showing consistent results for the ATS and AIT cases, and the transition from ATS to AIT by increasing the distance between the pillars is illustrated. The AIT phenomenon is shown to be periodic due to the periodicity of Fabry-Perot resonance. The band structures and the transmission spectra are calculated with the finite element method(FEM, COMSOL Multiphysics$^{\circledR}$).

\section{Unit cell model}
\begin{figure}[h]
	\centering
	\includegraphics[width=.8\linewidth]{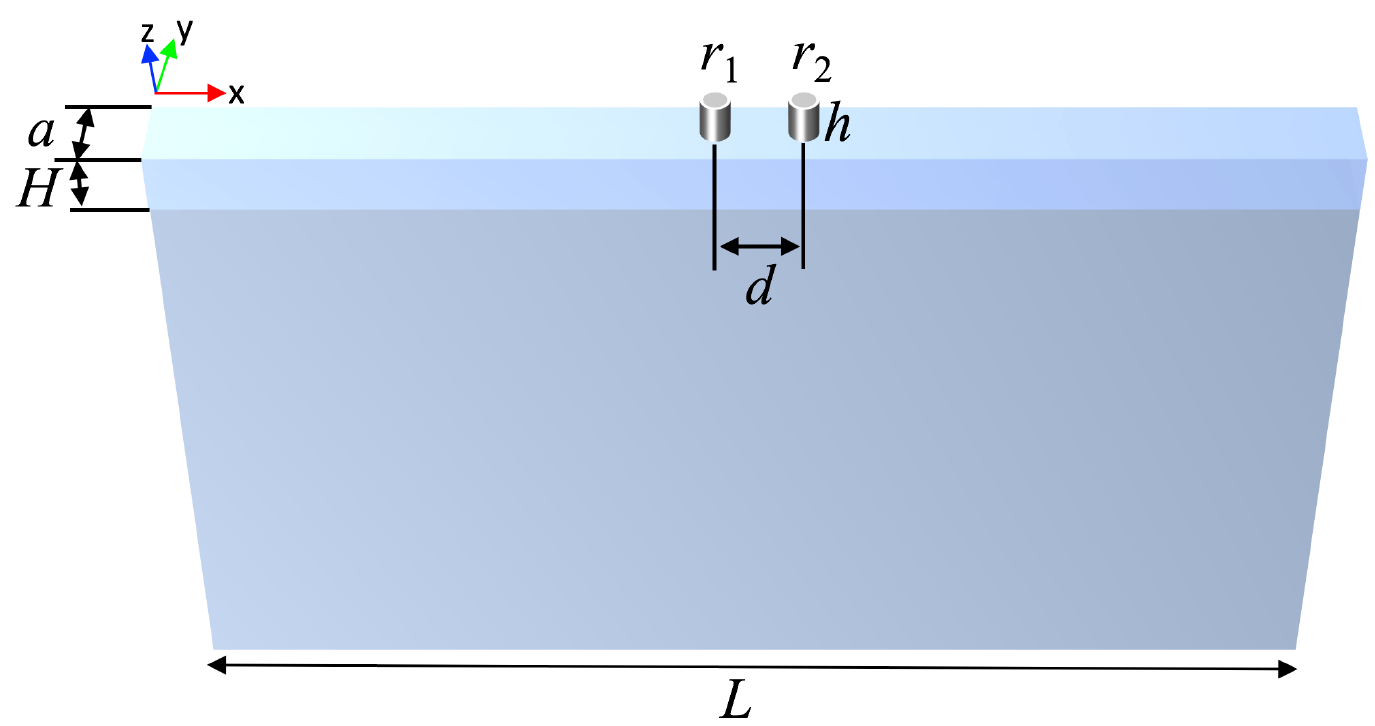}
	\caption{Unit cell of two lines of cylindrical Ni pillars on the SiO$_{2}$ film deposited on a 90ST-cut quartz. The period is $a$ along the $y$ direction. Love waves propagate along the $x$-axis. $r_{1}$=$r_{2}$=0.2$a$, $h=0.6a$, $H$=2.4$\mu\textrm{m}$, $L$=$20a$, $a=2\mu\textrm{m}$.}
	\label{SCPl}
\end{figure}

The acoustic meta-surface is constituted of two pillar lines deposited on a bi-layered matrix, of which the unit cell is shown in Fig~\ref{SCPl}. The matrix is a $2.4\mu\textrm{m}$-height silica guiding layer ($\rho=2200$kg/m$^{3}$, $E=70\textrm{GPa}, \nu=0.17$) that covers a $40\mu\textrm{m}$-height 90ST-cut quartz substrate (Euler angles=(0\textdegree, 47.25\textdegree, 90\textdegree), LH 1978 IEEE) which has been rotated 90 degrees around the $z$-axis from the ST-cut quartz, to be able to generate fast SH waves ($c_{SH,sub}$= 5000$\textrm{m/s}$) by the electric field. 
The shear wave velocity in the silica film is 3438$\textrm{m/s}$, less than that in the 90ST-cut quartz substrate, indicating the existence of Love waves. This structure is chosen since it has been experimentally validated for Love waves \cite{liu_evidence_2014}. The guiding layer thickness is sub-wavelength since only the fundamental Love mode exists to avoid physical complexity.
The period along the $y$-axis is $a=2 \mu\textrm{m}$. Two cylindrical Ni pillars on the silica film have a radius of $r_{1}=r_{2}=0.2a$ and a height of $h=0.6a$. The distance between the centers of the two pillars is denoted by $d$. The inclusions are chosen because of their strong contrast in density and elastic constants with regard to the matrix. However, similar results can also be obtained by using other materials such as gold or diamond. 
The length of the unit cell $L$ is $20a$ to insure a decoupling between the unit cells. 
Floquet periodic boundary conditions are applied along the $x$ and $y$ directions. The bottom of the substrate is assumed fixed. Love waves propagate along the $x$-axis (i.e. the $y$-axis of the ST-cut quartz), where Rayleigh waves cannot be generated \cite{liu_design_2014} due to a zero electromechanical coupling factor in the substrate.
The surfaces of the pillars coincide with the plane $z=0$. The wavelength normalized energy depth (NED)\cite{liu_highly_2018} is calculated to select the surface modes:  
\begin{equation}
\textrm{NED}=\frac{\iiint_{\mathcal{D}}\frac{1}{2}T_{ij}S_{ij}^{*}(-z)dxdydz}{\lambda\iiint_{\mathcal{D}}\frac{1}{2}T_{ij}S_{ij}^{*}dxdydz} ,
\end{equation}
where $T_{ij}$ is the stress and $S_{ij}$ the strain. The star symbol (*) means the complex conjugate. $-z$ denotes that we integrate in the whole domain of the unit cell $\mathcal{D}$. $\lambda$ is the wavelength. 
This formula works well for a relatively large $k$ where the wave speed is less than the SH waves velocity of the substrate. For a relatively small $k$ (with wave speed greater than the SH waves velocity of the substrate), which corresponds to a large $\lambda$, we fix $\lambda$ to $2a$.
Surface modes have a NED $<2$. The NED can filter out the bulk modes as well as the plate modes that appear in our finite-depth substrate which is supposed to be semi-infinite for Love waves.

SAW include SH type SAW and Rayleigh type SAW. The SH ratio is calculated to distinguish the displacement components: 
\begin{equation}
\textrm{SH ratio}=\frac{\iiint_{\mathcal{D}}u_{SH}u_{SH}^{*}dxdydz}{\iiint_{\mathcal{D}}(u_{x}u_{x}^{*}+u_{y}u_{y}^{*}+u_{z}u_{z}^{*})dxdydz} ,
\end{equation}
where $u_{x}, u_{y}$ and $u_{z}$ are respectively the displacement along the $x, y, z$ directions. $u_{SH}$ is the SH displacement component that can be expressed as $u_{x}\cos\theta-u_{y}\sin\theta$, which is perpendicular to the wave vector $\bm{k}$. $\theta$ is the angle between $\bm{k}$ and the $y$-axis such that $\tan\theta=\frac{k_{x}}{k_{y}}$. In our case, $\bm{k}=\bm{k_{x}}$ and $u_{SH}=u_{y}$.

\section{resonant properties of single pillar line}
\begin{figure}[h!]
	\centering
	\includegraphics[width=\linewidth]{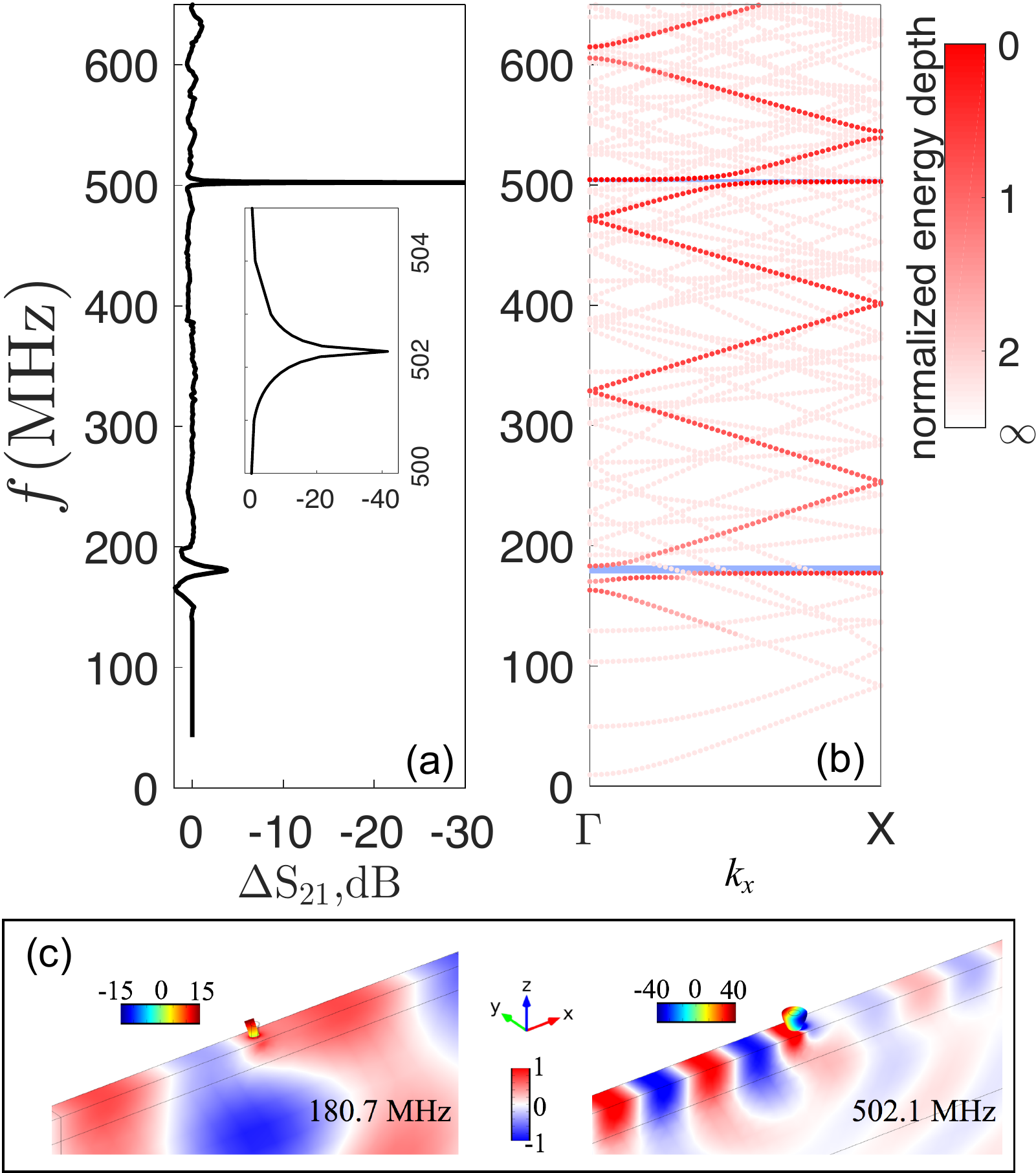}
	\caption{(a) Transmission spectrum of Love waves propagating through a single pillar line. Inset is the zoom of the torsional mode induced dip. (b)Band structure of SH modes in $\Gamma$X direction for the unit cell of a single pillar line. The red-white colors denote the normalized energy depth. A mode in red can be excited by Love waves. (c) $u_{y}$ component of the displacement fields for two local resonant pillar modes. The amplitudes in the pillar are normalized to the maximum amplitude in the SiO$_{2}$ film.  $r=0.2a$. $h=0.6a$, $a=2\mu\text{m}$.}
	\label{bsTr1Pl}
\end{figure}

Firstly, we study the resonant properties of the unit cell with only one pillar, which corresponds to a meta-surface with a single pillar line. The dispersion curves or band structure of SH modes (SH ratio$>$0.5) in the $x$ direction is shown in Fig~\ref{bsTr1Pl}(b). The X point is the irreducible Brillouin zone limit of the unit cell in the $x$ direction. The modes colors are determined by their NED values. The modes in red are well confined to the surface, and can therefore be excited by Love waves. Certain modes become pink as they are less confined to the surface. Two hybridization band gaps are observed respectively in the frequency range [177.3, 183.1] MHz and [501.8, 503.3] MHz, indicated in blue, which originates from the coupling of local resonant pillar modes and the SH SAW. Note that in the frequency range [100, 600] MHz, the wavelength ranges from 6.8 to 48.8$\mu m$. Therefore, our structure is indeed a meta-surface with sub-wavelength-scaled pillar size, periodicity and guiding layer thickness. 

\begin{figure}[]
	\centering
	\includegraphics[width=\linewidth]{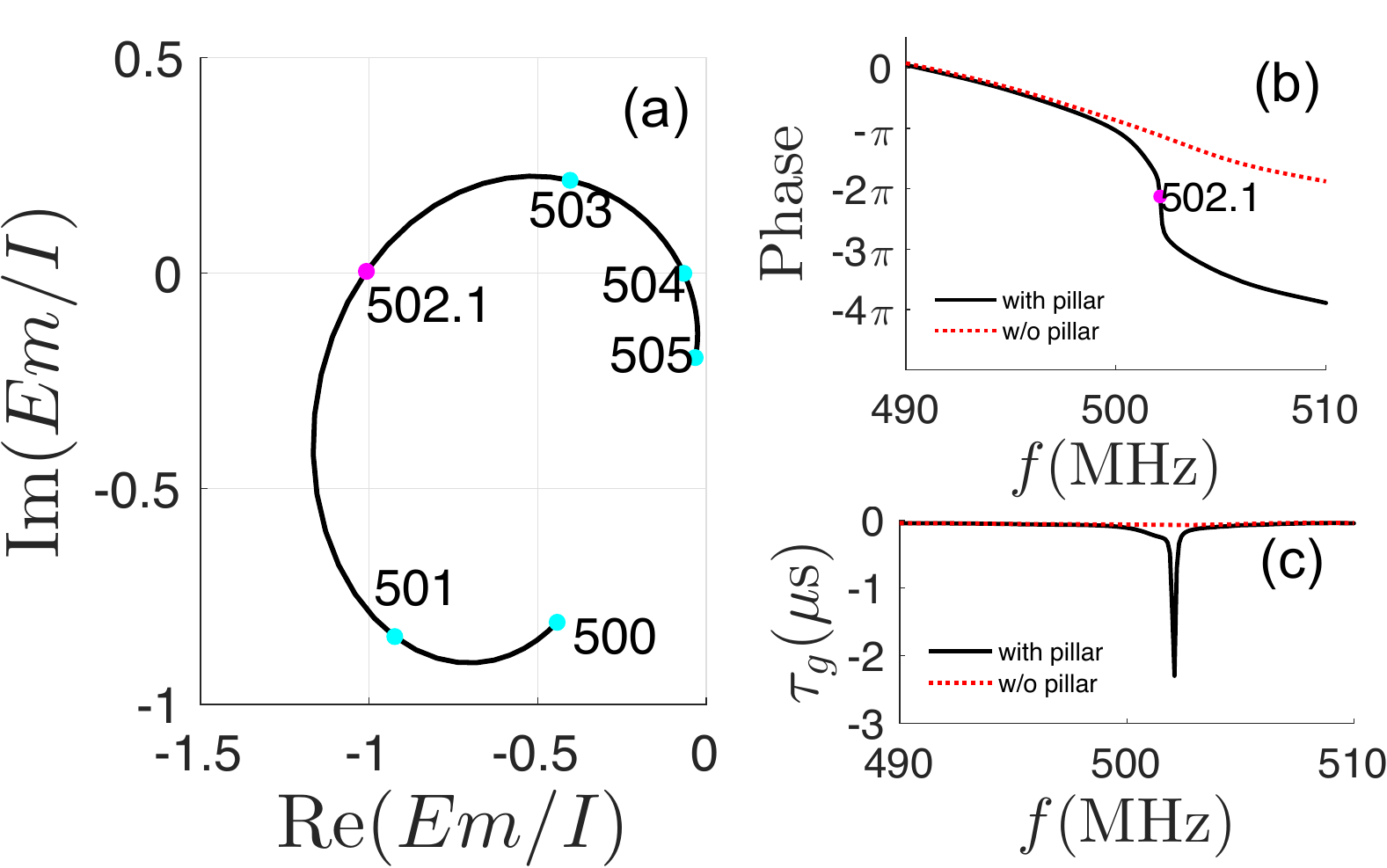}
	\caption{(a) Complex plot of the normalized emitted waves in the frequency range [500, 505] MHz. The rose dot corresponds to the transmission dip at $502.1$ MHz. (b) Phase of total transmitted waves. Red dotted line denotes the reference phase (incident waves without pillar). (c) Group delay time of the transmitted waves with and without pillar.}
	\label{emi1Pl}
\end{figure}
\begin{figure*}[]
	\centering
	\includegraphics[width=\linewidth]{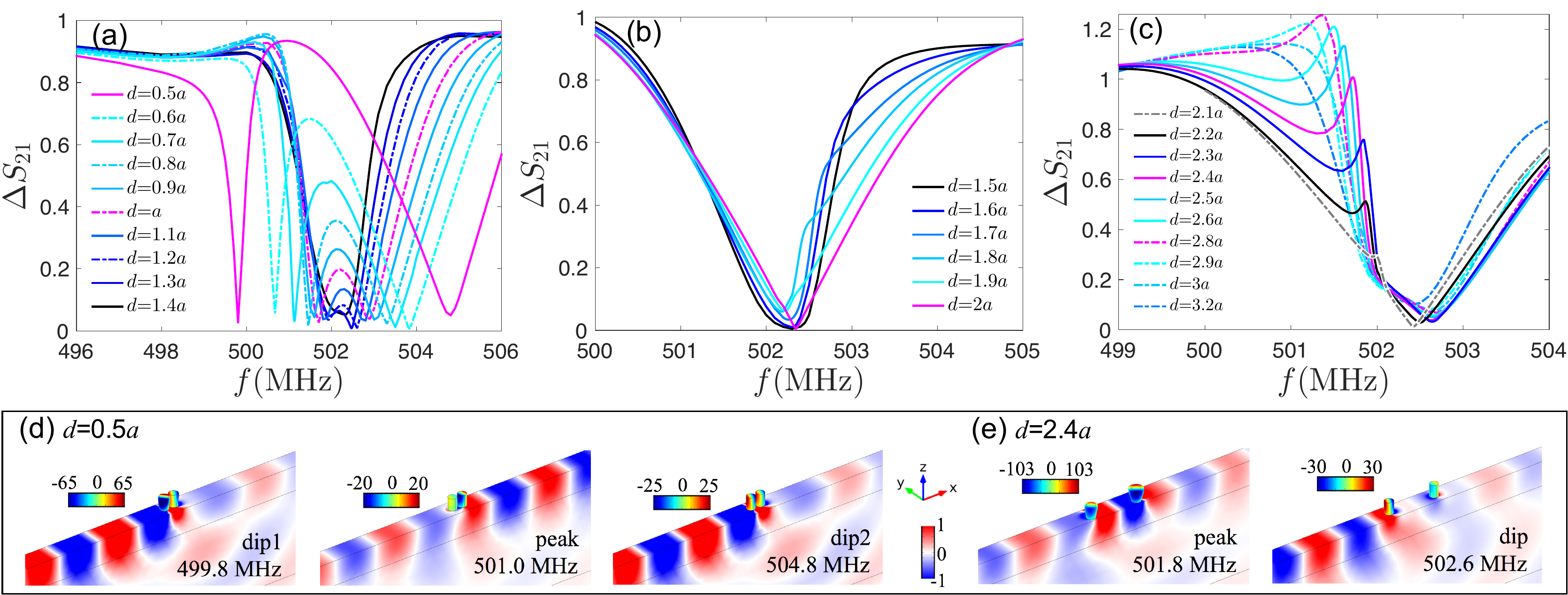}
	\caption{Transmission spectrum of Love waves propagating through the two identical pillar lines around the resonant frequency of torsional mode at 502.1 MHz with different central distance $d$: (a) presents the lifting of degeneracy and the apparition of Autler-Townes Splitting with the decrease of $d$; (b) shows the red-shift of cavity mode by increasing $d$ when $d<2a$ (c) presents the appearance of cavity mode peak when $d>2a$. (d) and (e) are the displacement fields $u_{y}$ at the dips and peaks for $d$=$0.5a$ and $2.4a$, respectively. $r_{1}$=$r_{2}$=0.2$a$, $a=2\mu\textrm{m}$. }
	\label{Tr-ed}
\end{figure*}

The transmission spectra is calculated by simulating the same dispersive SAW device introduced in our preceding works \cite{liu_highly_2018,yankin_finite_2014}, which has been validated by experiment. The model consists of two parts of aluminum inter-digital transducers (IDTs) with the unit cell located between IDTs. This device model is periodic along the $y$ direction. Other lateral sides and the bottom are assumed fixed. 
The input IDT is given a $V_{0}=1V$ harmonic voltage signal. 
The output is measured by averaging the voltage difference between odd and even electrodes. Dissipation is neglected in the results shown in the manuscript. However, its effect is discussed in Appendix A supporting the validity of the presented features.
The width of the electrodes of IDTs $L_{IDT}$ is updated for every frequency in the spectrum according to the relation $L_{IDT}=\frac{\lambda}{4}=\frac{v}{4f}$. $v$ is the velocity of Love waves for $H=2.4\mu m$, resulting from the basic dispersion relation of Love waves. That is, each frequency corresponds to a single wavelength. 
The frequency responses are then normalized by that of the matrix (without pillar), referred to as relative transmissions $\Delta S_{21}$. Fig~\ref{bsTr1Pl}(a) shows the transmission spectrum of the single pillar line. It can be seen that the transmission spectrum corroborates well with the band structure prediction. The displacement fields and the deformations at the two dips are shown in Fig~\ref{bsTr1Pl}(c). Due to their large SH component ratio, as well as the exclusive generation of SH waves by the electric field, we only show the transverse component $u_{y}$. The amplitude in pillar is normalized to that in the matrix and is indicated beside the pillar. The two dips correspond to the pillar's intrinsic bending and torsional modes, respectively. The torsional mode induced transmission dip is better attenuated, since the excited torsional motion leads to, on the side opposite to the incident wave, a wave of identical amplitude and opposite phase, which is responsible for a destructive interference.   
To further confirm this mechanism, the emitted wave $Em$ is calculated by subtracting the incident waves $I$ (transmitted waves on the bare matrix without pillar) from the totally transmitted waves $T$: $Em=T-I$. $Em$ is then normalized by $I$. Fig~\ref{emi1Pl}(a) is the complex plot of the normalized emitted wave in the frequency range [500, 505] MHz around the torsional mode. The rose dot corresponds to the dip at resonant frequency of $502.1$ MHz. This mode falls at point (-1,0), which refers to the same amplitude with a phase shift of 180\textdegree  with respect to the incident waves. This results in a destructive interference and a strong attenuation in transmission.
The phase shift of full transmitted waves is shown in Fig~\ref{emi1Pl}(b). Red dotted lines represent the incident waves.  The dip corresponds to a $\pi$ change in phase with respect to the incident waves. The phase shift is $0$ ($2\pi$) before (after) the resonant frequency, meaning that the transmitted waves are in phase with the incident waves when they deviate from the resonant frequency. 
The corresponding phase derivative or group delay time $\tau_{g}$ of the transmitted waves is shown in Fig~\ref{emi1Pl}(c), with $\tau_{g}=\frac{d\phi}{d\omega}$, where $\phi$ denotes the phase and $\omega=2\pi f$ is the angular frequency. The waves are delayed at the resonant frequency. It can be seen that the resonance is characterized by a rapid variation of phase and amplitude.

By scaling the model in the same proportion, 
we can obtain the same dimensionless frequency 
$\omega a/c_{SH,sub}$ 
of a mode.
Since the pillar intrinsic mode is almost independent of the periodicity, one can also tune the pillar mode frequency only by modifying the pillar size. Note that this frequency needs to be within the allowable operating range of the materials.

\section{Two lines of identical pillars: Autler-Townes Splitting \& cavity mode}\label{sect.LR}
\begin{figure}[h!]
	\centering
	\includegraphics[width=.8\linewidth]{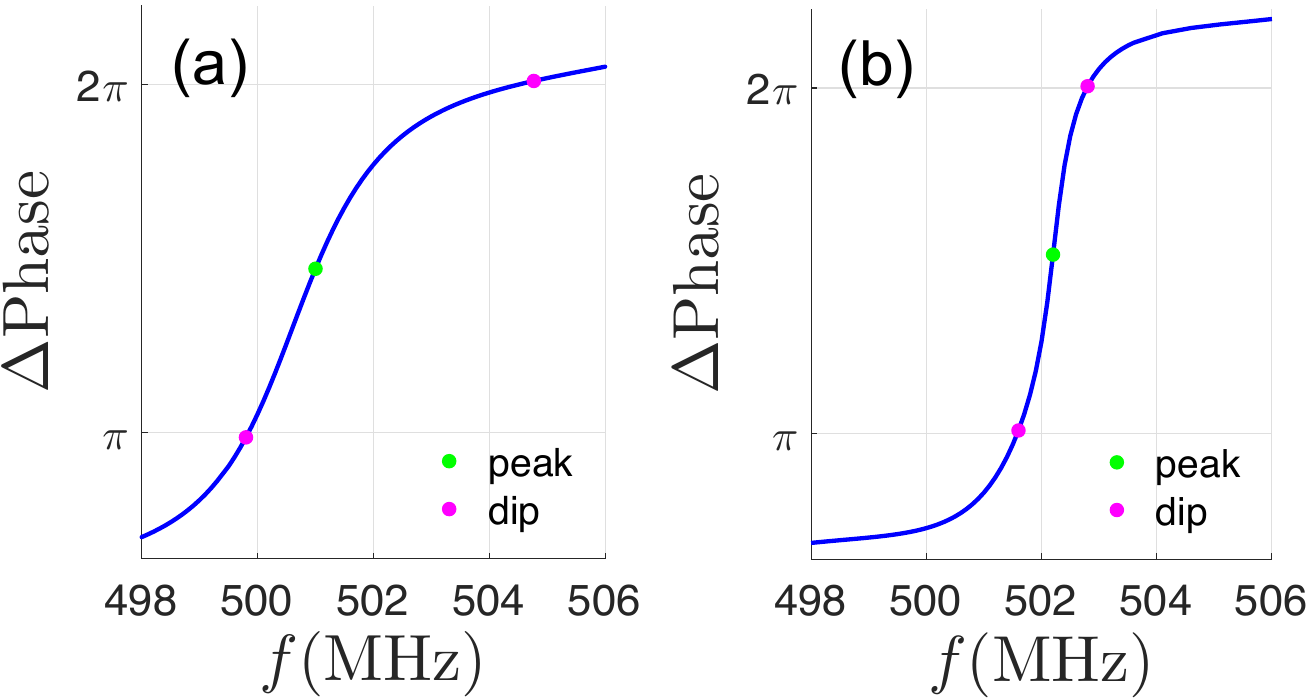}
	\caption{Pillar phase differences for ATS in the case of (a) $d=0.5a$ and (b) $d=a$. Rose and green dots correspond to transmission dips and peaks, respectively.  $r_{1}$=$r_{2}$=0.2$a$, $a=2\mu\text{m}$.}
	\label{ATSpl}
\end{figure}

Transmission spectra are then calculated around the torsional mode frequency for two lines of identical pillars. In addition to our intuitively predicted transmission dip, different phenomena appear when we gradually increase the distance $d$ between the pillar lines.
When $d<1.4a$ (Fig~\ref{Tr-ed}(a)), a coupling effect arises between the pillar lines, causing a lifting of frequency degeneracy of the pillar torsional mode, and the original transmission dip splits into two dips with a transparency window in the middle, referred to as acoustic analogue of Autler-Townes Splitting (ATS). The coupling becomes stronger when $d$ gets smaller, as shown in Fig~\ref{Tr-ed}(a). Note that the pillar coupling also depends on the pillar mass. Therefore, the distance limit $1.4a$ can be detuned by changing the pillar size (radius or height). The displacement fields $u_{y}$ at the dips and peak for $d=0.5a$ are shown aside in  Fig~\ref{Tr-ed}(d). It is found that the largest amplitude is located in the pillars for dip1 at 499.8 MHz. The two pillars are in opposite phase at dip1 frequency and in-phase at dip2 frequency, which is a feature of the ATS resonance that can be confirmed by calculating the phase difference between the two pillars. 
Since the pillars are in torsional mode in the range of measurement, all the points on the side of the pillar that faces the incident waves are in phase. The phases of $u_{y}$ on the wave-facing sides of the two pillars are probed. The cases of $d=0.5a$ and $d=a$ are shown as examples in Fig~\ref{ATSpl}(a) and (b). Rose and green dots correspond to transmission dips and peaks, respectively. It can be seen that the two pillars have a phase difference of $\pi$ at the first dip, meaning that they are 180\textdegree out-of-phase. The phase difference is $2\pi$ at the second dip, indicating that they are in-phase. This reveals that the pillar vibrations are symmetrical at the dip1 frequency and asymmetrical at the dip2 frequency. We think this behavior leads to different quality factors of the two dips.  

\begin{figure}[h!]
	\centering
	\includegraphics[width=\linewidth]{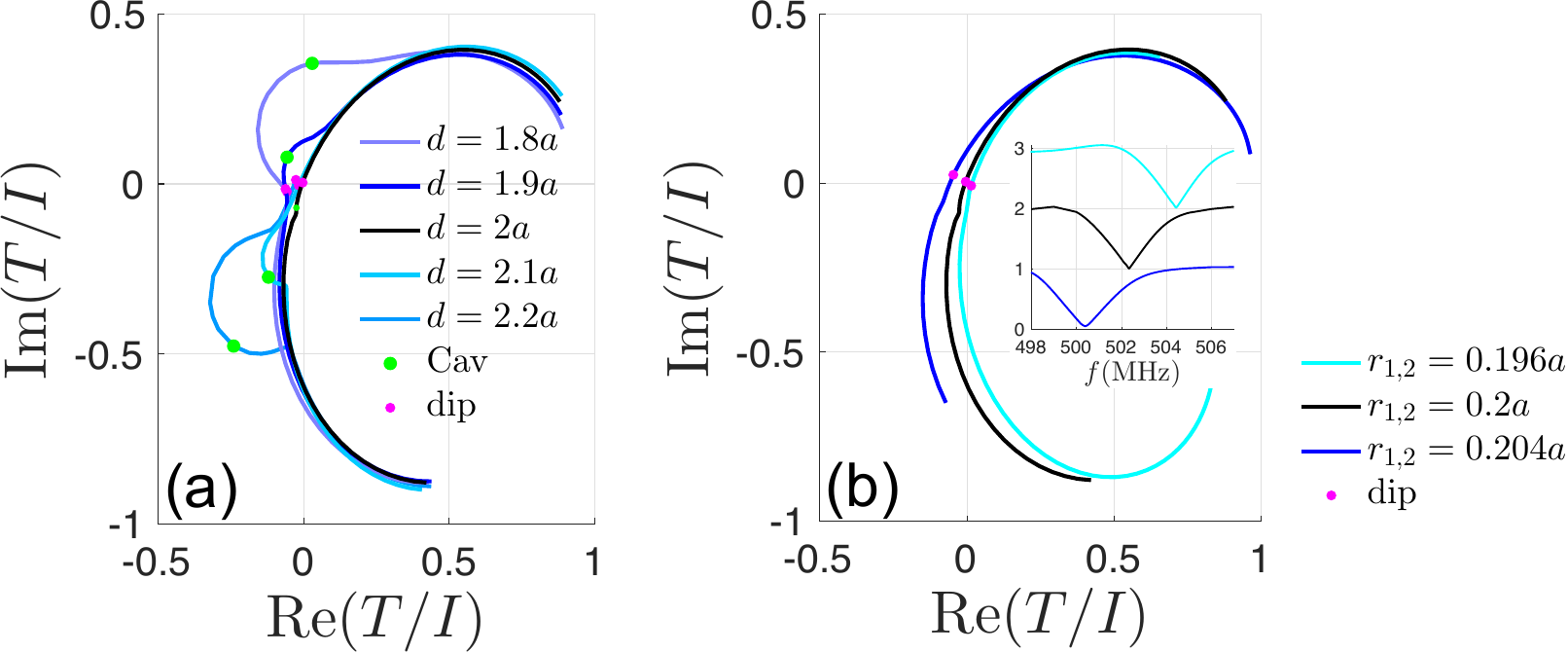}
	\caption{(a)Complex plots of the normalized transmissiosn in the frequency range [500, 505] MHz for $d$ varies from $1.8a$ to $2.2a$. The rose and green marks correspond to the dip and cavity mode frequencies, respectively. $r_{1}$=$r_{2}$=0.2$a$. 
		(b) Complex plots of the normalized transmissions for $d=2a\approx\lambda/2$ when all the pillars' radius vary from $0.196a$ to $0.204a$. Inset shows the transmission spectra of corresponding curves.
	}
	\label{FPtrans}
\end{figure}
\begin{figure}[h!]
	\centering
	\includegraphics[width=\linewidth]{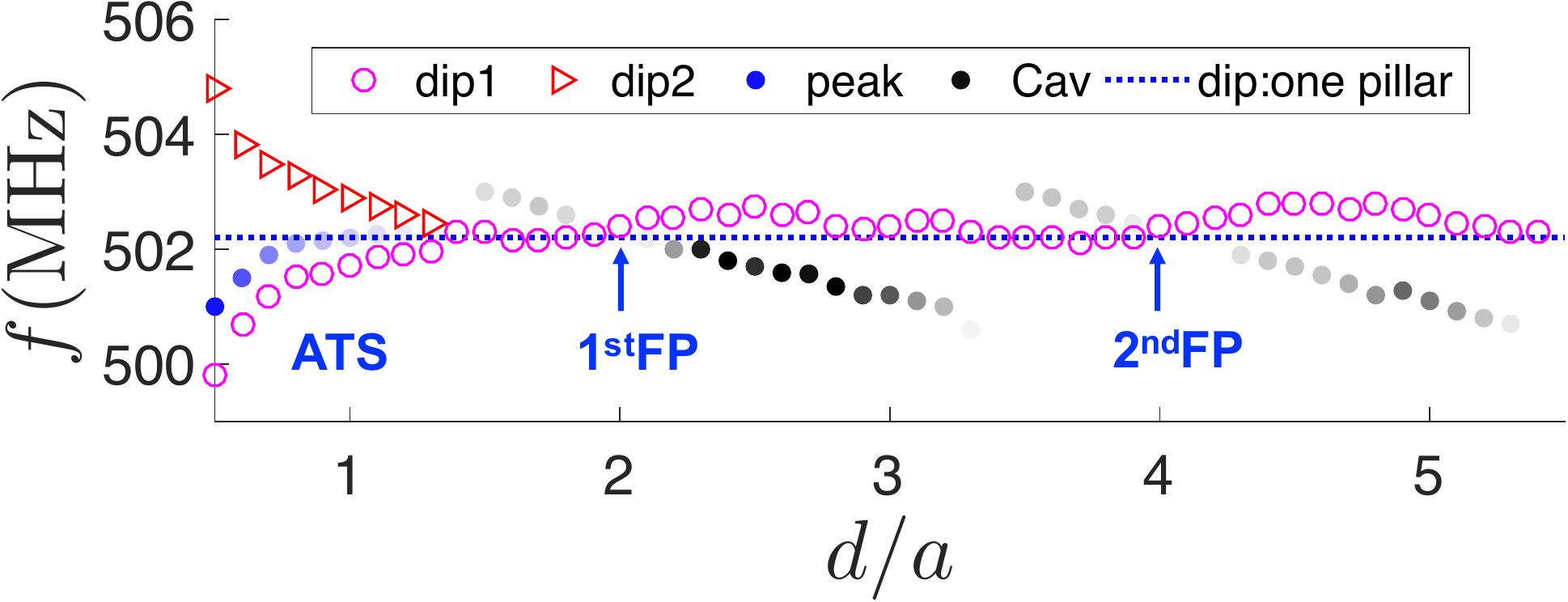}
	\caption{Dip, peak and cavity mode frequencies as functions of the distance $d$ between two identical pillar lines. Blue dotted line is the dip frequency of a single pillar line. ATS appears in the coupling region of $d<1.4a$. The $1^{st}$ and $2^{nd}$ Fabry-Perot resonances fall at $d$=$\lambda/2$ and $\lambda$, respectively. $r_{1}$=$r_{2}$=0.2$a$, $a=2\mu\textrm{m}$.}
	\label{f-ec}
\end{figure}
\begin{figure*}[t]
	\centering
	\includegraphics[width=\linewidth]{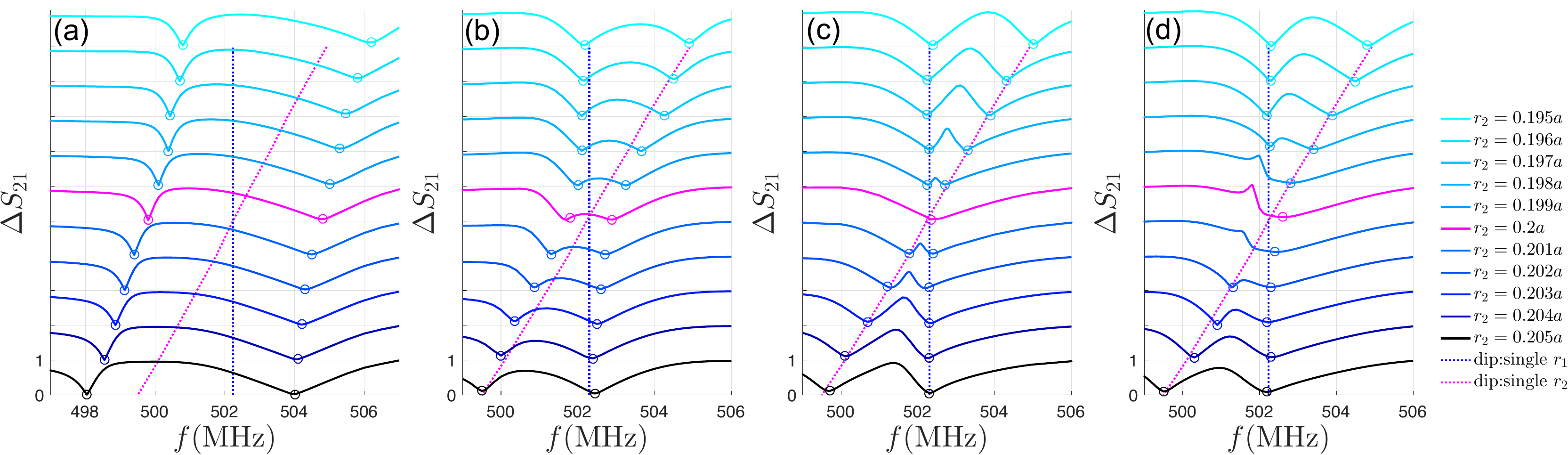}
	\caption{Transmission spectra of Love waves for $d$= (a)$0.5a$, (b)$a$, (c)$2a$, (d)$2.4a$ when gradually changing the radius of the second pillar $r_{2}$ from $0.195a$ to $0.205a$. The transmission spectra associated with the different values of $r_{2}$ have been shifted vertically with respect to each other by one unit. The radius of the first pillar $r_{1}$ is fixed to $0.2a$. $a=2\mu\text{m}$.}
	\label{cpFP}
\end{figure*}

Since we work around 500MHz with Love waves velocity $v$ around 4200 m/s, the wavelength $\lambda=v/f$ is therefore around 8.4$\mu$m.  $d\approx2a$ ($a=2\mu m$) indicates a distance around $\lambda/2$ corresponding to the Fabry-Perot (FP) resonance. This resonance is almost invisible in the transmission spectrum when it matches the pillar torsion mode, since the waves are almost totally reflected. However, when we change the distance between pillars around $\lambda/2$, the proximity of FP resonance with the pillar mode gives rise to the cavity modes at the two edges of the dip, as shown in Fig~\ref{Tr-ed}(b) and (c). In these cases, the two pillars act like partial reflectors, and the normally incident waves are multiply reflected to produce multiple transmitted waves with path difference equal to $n\lambda$, where $n$ is an integer. In this way, constructive interference occurs, leading to a resonant enhancement. The two pillars along with the guiding layer between them become a cavity to confine the waves. Nevertheless, in the transmission spectra, the behaviors for $d<2a$ (Fig~\ref{Tr-ed}(b)) are less marked than that for $d>2a$ (Fig~\ref{Tr-ed}(c)), where peaks rise at the lower edge of the dip, and give rise to Fano-like resonance line-shapes. However, we have verified that it can not be fitted by a Fano type formula. 
The displacement field $u_{y}$ at the peak and dip for $d=2.4a$ are shown aside in Fig~\ref{Tr-ed}(e), since the transmission curve for $d=2.4a$ presents the most confined peak. At the peak frequency, large amplitudes are observed for the three parts of the cavity (two pillars and the guiding layer in between). This behavior is different from that of ATS, the largest amplitude in the pillars occurs at the peak where the waves in the guiding layer are highly confined in the cavity.

To show more clearly the transition of the cavity mode with respect to the dip when $d$ changes, we draw the complex plots of the normalized transmissions in Fig~\ref{FPtrans}(a). The cavity mode is manifested as an additional perturbation on the original ellipse. Since the phase changes clockwise, it can be seen that the cavity mode passes the dip as $d$ increases. When the cavity mode approaches to the dip, the perturbation decreases. For $d=2a$, the cavity mode coincides with the FP resonance and becomes almost invisible. Additionally, we can see that the behaviors for $d<2a$ and for $d>2a$ are quite similar, with the perturbation frequency either larger or smaller than the dip frequency.


Fig~\ref{FPtrans}(b) shows the complex plots of the normalized transmissions of Love waves for $d=2a$ when we change the radius of both pillars from $0.196a$ to $0.204a$, indicating a shift in pillar resonant frequency. It is found that the cavity mode remains almost invisible, i.e. still coincides with the FP resonance. That is because the distance between pillars corresponding to the FP resonance is almost unchanged in our range of measurement (from 498 to 506 MHz), i.e. $\lambda/2$ is always around $d=2a$. Therefore, when we change the pillar vibration frequency, we obtain the parallel cavity modes for the same value of $d$.    

To give an overview of the resonance behaviors, the dip and peak frequencies for different $d$ are shown in Fig~\ref{f-ec}. Blue dotted line is the dip frequency of the single pillar line. It can be seen that the pillar coupling induced ATS is in the region $d<1.4a$, where the two dips are mismatched with the single pillar resonant frequency. This coupling disappears when $d$ exceeds $1.4a$, and only one dip (quasi-zero transmission) remains. This dip matches the single pillar resonant frequency except when the cavity modes appear below (upon), the dip frequency shifts slightly upward (downward). The interaction between the pillars is much stronger for $d/a$=2 than that for $d/a$=4. Therefore, the resonances around $d$=$\lambda$ are too weak to be observed. Note that the frequency of the cavity modes changes much slower than that of the FP resonance. The $1^{st}$ and $2^{nd}$ FP resonance exists only in the very closed regions around $d=2a$ and $4a$, respectively. In these two cases, FP resonances are particular cases of the cavity modes when the later coincide with the dip. In the other regions, one should avoid to mix up the cavity mode with the FP resonance.

\section{Two lines of dissimilar pillars: Autler-Townes Splitting, cavity \& Acoustically Induced Transparency}
Since an increase of pillar radius or height will induce a decrease in the torsional mode frequency, one can gradually tune the position of the dip by modifying the pillar size. Here we modify the pillar radius as example since the radius is easier to be tuned in experimental process.

\begin{figure}[h!]
	\centering
	\includegraphics[width=.6\linewidth]{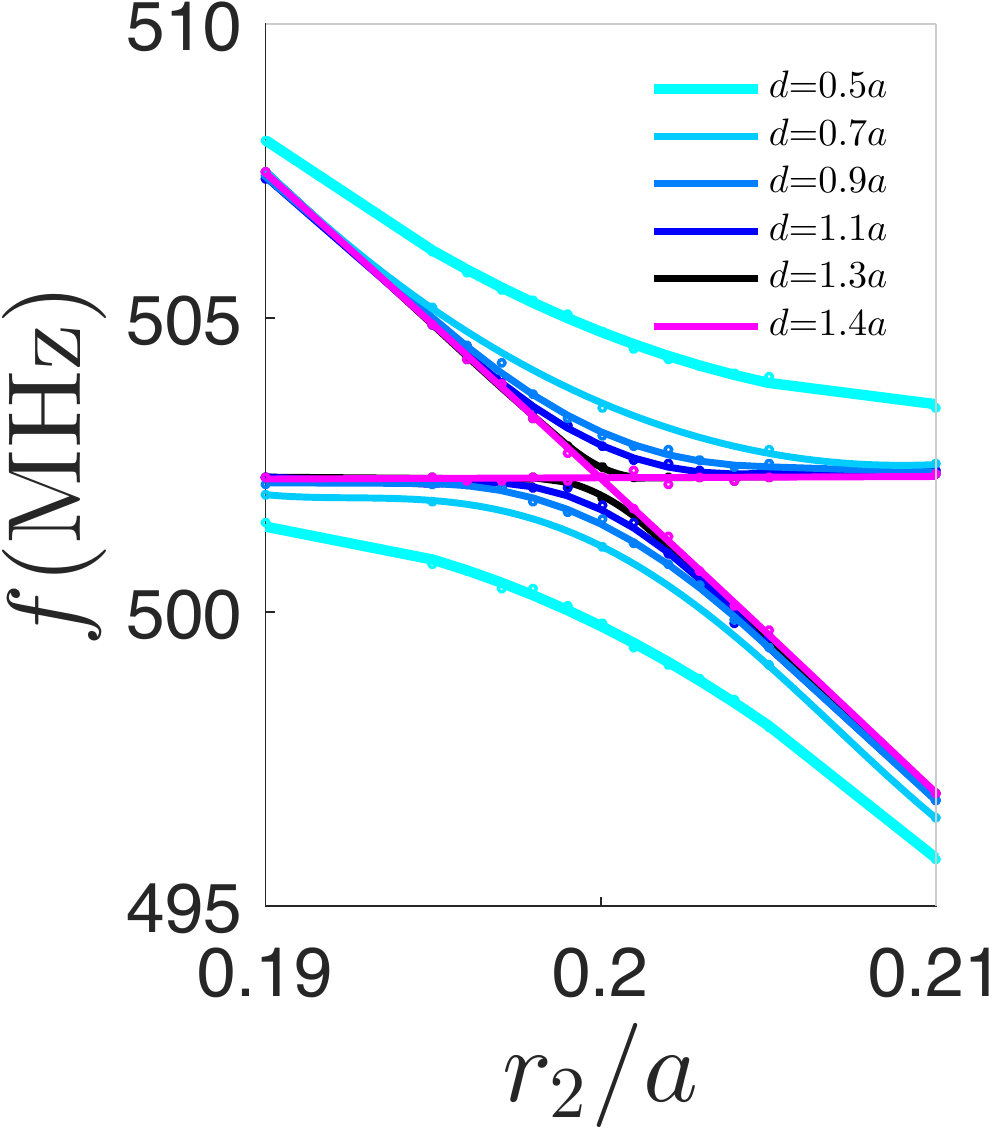}
	\caption{Anti-crossing lines for ATS when $d<1.4a$. Dip frequencies for different $d$ as a function of the second pillar radius $r_{2}$. $r_{1}$ is fixed to $0.2a$. $a=2\mu\text{m}$.}
	\label{cross}
\end{figure}

In the case below, the second pillar radius $r_{2}$ is tuned from $0.195a$ to $0.205a$, while the first pillar radius $r_{1}$ being fixed to $0.2a$.  
Fig~\ref{cpFP} shows the transmission spectra of Love waves propagating through the two dissimilar pillar lines for different $d$ that remains unchanged for each case. Let un notice that the curves for different $r_{2}$ are shifted vertically with respect to each other. Dotted blue and rose lines denote the dip positions for a single line of pillars with radius equals to $r_{1}$ and $r_{2}$, respectively. Fig~\ref{cpFP}(a) corresponds to the case $d=0.5a$, where the coupling between the pillars is so strong that the two dips stay all the way mismatched with their corresponding single pillar dip positions.  
When $d$ gets larger, the coupling becomes weaker: Fig~\ref{cpFP}(b) corresponds to the case $d=a$. It is found that this coupling decreases with the increase of radius mismatch. In the case of $r_{2}=0.195a$ and $0.205a$, each dip almost coincides with the corresponding resonant frequency of one single pillar. 
In order to show the anti-crossing lines of ATS for $d<1.4a$, we plotted in Fig~\ref{cross} the dip frequencies for different $d$ as functions of $r_{2}$. It can be seen that with the increase of pillar distance, the anti-crossing lines get closer to the crossing line (for $d=1.4a$), and that each anti-crossing line will rejoin its individual pillar resonant frequency when the radius mismatch is sufficiently large. This means the pillar coupling effect decreases when their distance or/and their radius difference increases.

\begin{figure}[]
	\centering
	\includegraphics[width=\linewidth]{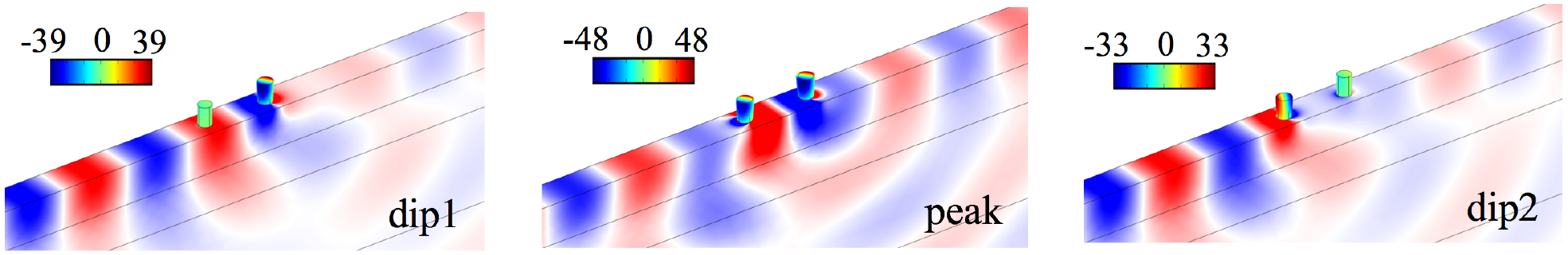}
	\caption{Displacement field $u_{y}$ at the dips and peak for AIT in the case of $d=2a$. The two pillars differ in radius with $r_{2}=0.202a$ and $r_{1}=0.2a$. $a=2\mu\text{m}$.}
	\label{dispAIT}
\end{figure}

\begin{figure}[t]
	\centering
	\includegraphics[width=\linewidth]{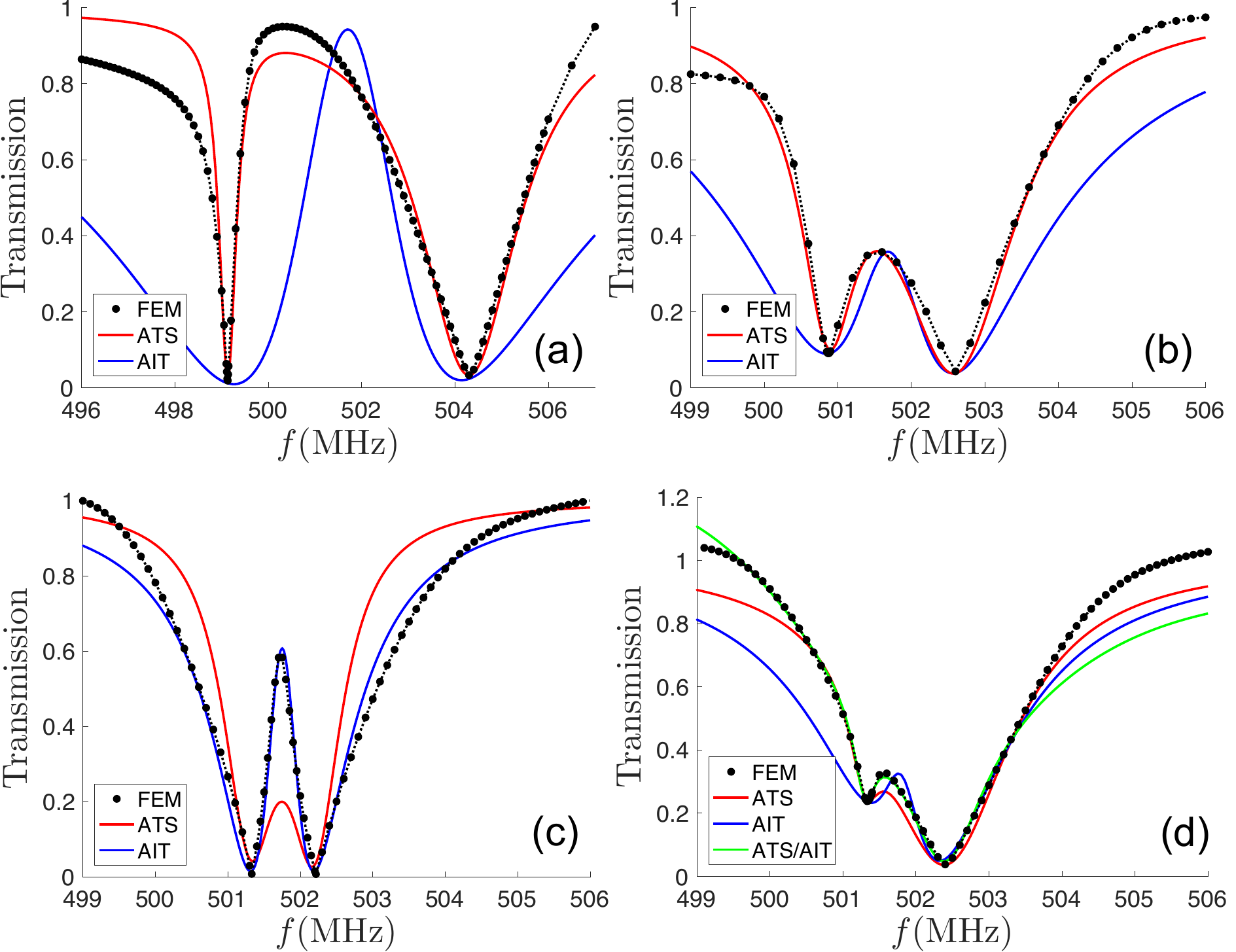}
	\caption{Transmission spectra and model fits of ATS and AIT, for $r_{1}=0.2a$, $r_{2}=0.202a$ and $a=2\mu\text{m}$. Numerical data (black dots) are presented together with the best fits of functions $T_{ATS}$
		(red lines) and $T_{AIT}$
		(blue lines). For (a) $d$=$0.5a$ and (b)$d=a$, $T_{ATS}$ fits the numerical data better than $T_{AIT}$. (c) For $d$=$2a$, $T_{AIT}$ fits the numerical data better than $T_{ATS}$. (d) For $d$=2.4$a$, $T_{ATS/AIT}$ (green line) can be used to fit the numerical data whereas $T_{ATS}$ and $T_{AIT}$ do not fit well.}
	\label{fit}
\end{figure}

For $d=2a$ as shown in Fig~\ref{cpFP}(c), two pillars with different radius give rise to two dips with a transparency window in the middle. These peaks have a narrower line-shapes compared with the cases of ATS (see also Fig~\ref{fit}). Each dip is consistent with the corresponding dip frequency of one single pillar since no more coupling exists, which means each of them originates from individual pillar’s torsional mode. 
The displacement fields $u_{y}$ for $r_{2}=0.202a$ are presented in Fig~\ref{dispAIT}. It can be seen that each dip corresponds to a large amplitude in a single pillar, indicating the attenuation of transmission at each pillar's resonant frequency due to the destructive interferences.
The peak correspond to a Fabry-Perot resonance since $d$ is close to $\lambda/2$. The two detuned pillar modes act as two partial reflectors and are
hence able to support the constructive interference of the FP resonance, where large amplitudes are observed for the three parts of the cavity (two pillars and the guiding layer in between).   
This three-resonance system induced transparency window is referred to as the Acoustic analogue of Electromagnetically Induced Transparency, also called Acoustically Induced Transparency (AIT). 
The peak rises and gets wider with the increase of radius difference. 
As for $d=2.4a$ shown in Fig~\ref{cpFP}(d), where a cavity mode peak is observed for the two identical pillars, it is found that with the increase of pillar radius mismatch, the peak confinement decreases and the dip1 becomes evident for $r_{2}=0.198a$ and $0.202a$.
Compared with the case of $d=2a$ (Fig~\ref{cpFP}(d)), the peak between two dips is less confined due to the red shift of cavity mode with respect to the pillar resonant frequency.
Note that AIT requires a well excited resonance between the two dips. Therefore, in the case of $d=2.4a$, the transparency window is only two dips resulting from the different resonant frequencies of two pillars.

The transmission spectra of AIT are similar with those of ATS, however, they originate from different mechanisms. ATS appears only when the two pillars are coupled to each other, and exists even when the two pillars are identical. AIT appears when $d$ is out of the pillar coupling region and only when the two pillars are different. The pillars and the cavity interact at the peak. Moreover, AIT requires a clearly identified 3-level resonant system, which is not the case for ATS.

Besides the different mechanisms related to ATS and AIT as presented above, corresponding analytical formulas of ATS and AIT for transmission spectra can be used to fit the numerical data to better distinguish these different transparency windows. The transmission curves for ATS can be written as the sum of two separate inverse Lorentzian profiles representing the two dips, while the transmission for AIT is expressed as the difference of a broad Lorentzian profiles and a narrow one with a similar central frequency \cite{peng_what_2014,liu_experimental_2016}:

\begin{equation}
T_{ATS}=1-\frac{C_{1}(\Gamma_{1}/2)^2}{(f-\delta_{1})^2+(\Gamma_{1}/2)^2}-\frac{C_{2}(\Gamma_{2}/2)^2}{(f-\delta_{2})^2+(\Gamma_{2}/2)^2},
\end{equation}
\begin{equation}
T_{AIT}=1-\frac{C_{+}(\Gamma_{+}/2)^2}{(f-\delta_{c}-\epsilon)^2+(\Gamma_{+}/2)^2}+\frac{C_{-}(\Gamma_{-}/2)^2}{(f-\delta_{c})^2+(\Gamma_{-}/2)^2},
\end{equation}
where $C_{1}$, $C_{2}$, $C_{+}$, $C_{-}$ are the amplitudes of the Lorentzian profiles, $\Gamma_{1}$, $\Gamma_{2}$, $\Gamma_{+}$, $\Gamma_{-}$ are their full width at half maximum (FWHM). $\delta_{1}$, $\delta_{2}$, $\delta_{c}$ are the central frequencies with $\epsilon$ denoting a possible slight shift on $\delta_{c}$. $\Gamma_{1}$, $\Gamma_{2}$, $\delta_{1}$, $\delta_{2}$ and $\delta_{c}$ can be directly taken from the transmission spectra.

In an intermediate state, the transmission spectra can be fitted by a transition formula that considers both the features of ATS and AIT:

\begin{equation}
\begin{split}
T_{ATS/AIT}=1-\frac{C_{a}(\Gamma_{a}/2)^2}{(f-\delta_{1})^2+(\Gamma_{a}/2)^2}-\frac{C_{b}(\Gamma_{b}/2)^2}{(f-\delta_{2})^2+(\Gamma_{b}/2)^2}\\
-\frac{(f-\delta_{1})C_{d}(\Gamma_{d}/2)^2}{(f-\delta_{1})^2+(\Gamma_{d}/2)^2}+\frac{(f-\delta_{2})C_{e}(\Gamma_{e}/2)^2}{(f-\delta_{2})^2+(\Gamma_{e}/2)^2},
\end{split}
\end{equation}
where $C_{a}$, $C_{b}$, $C_{d}$, $C_{e}$, $\Gamma_{a}$, $\Gamma_{b}$, $\Gamma_{d}$, $\Gamma_{e}$ are parameters to be determined. Note that this formula can also be used to fit the ATS 
and AIT 
cases.

Fig~\ref{fit} shows the numerical data (black dots) together with the best fit functions $T_{ATS}$ (red lines) and $T_{AIT}$ (blue lines) for the four values of $d$ in Fig~\ref{cpFP} which stand for strong coupling ATS, weak coupling ATS, AIT and intermediate states, respectively, in the case of $r_{2}=0.202a$ and $r_{1}$ fixed to $0.2a$. 
The best fit functions are determined by resorting to the 
least-squares method, that is, by searching the fit parameters that minimize the sum of 
squared errors between numerical data and fit function: 
\begin{equation}
\sum (FEM_{i}-F_{i})^2, i\in[1,N]
\end{equation}
where $N$ is the number of values calculated by FEM. $F_{i}$ is the value of $T_{ATS}$ or $T_{AIT}$ corresponding to each data frequency. 
It can be seen that, as expected, for $d=0.5a$ and $a$ where coupling exists between the pillars, $T_{ATS}$ fits the numerical data much better than $T_{AIT}$. Whereas for $d=2a$ 
, $T_{AIT}$ fits the numerical data better than $T_{ATS}$. For $d=2.4a$, both $T_{ATS}$ and $T_{AIT}$ do not fit the numerical data well. In this case, $T_{ATS/AIT}$ (green line) fits the data better.

\begin{figure}[t]
	\centering
	\includegraphics[width=\linewidth]{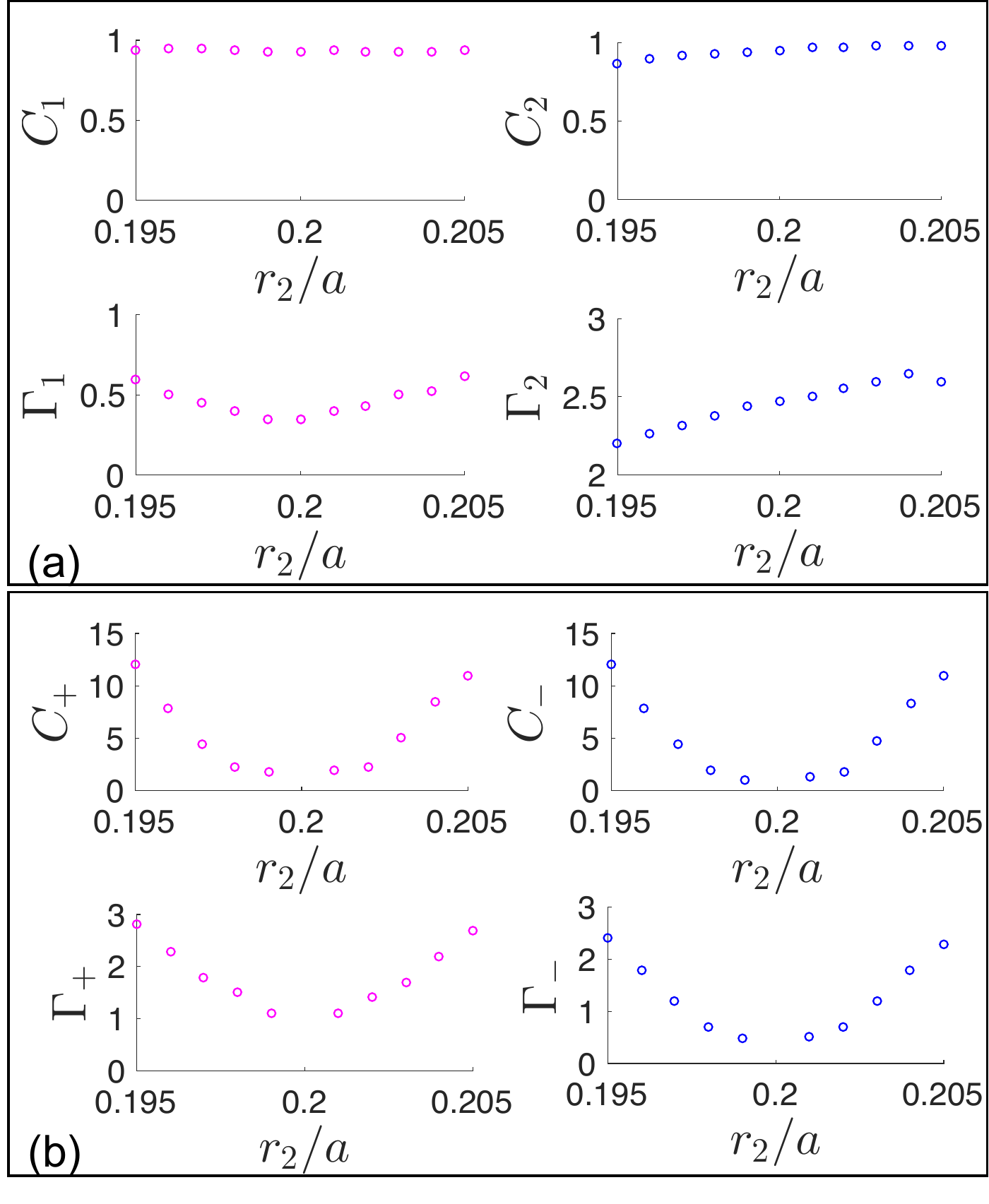}
	\caption{Fit parameters as functions of $r_{2}/a$ for (a)$d=0.5a$ with model fits $T_{ATS}$; (b)$d=2a$ with model fits $T_{AIT}$. $r_{1}=0.2a$, $a=2\mu\text{m}$}
	\label{fitpara}
\end{figure}
\begin{figure}[]
	\centering
	\includegraphics[width=.9\linewidth]{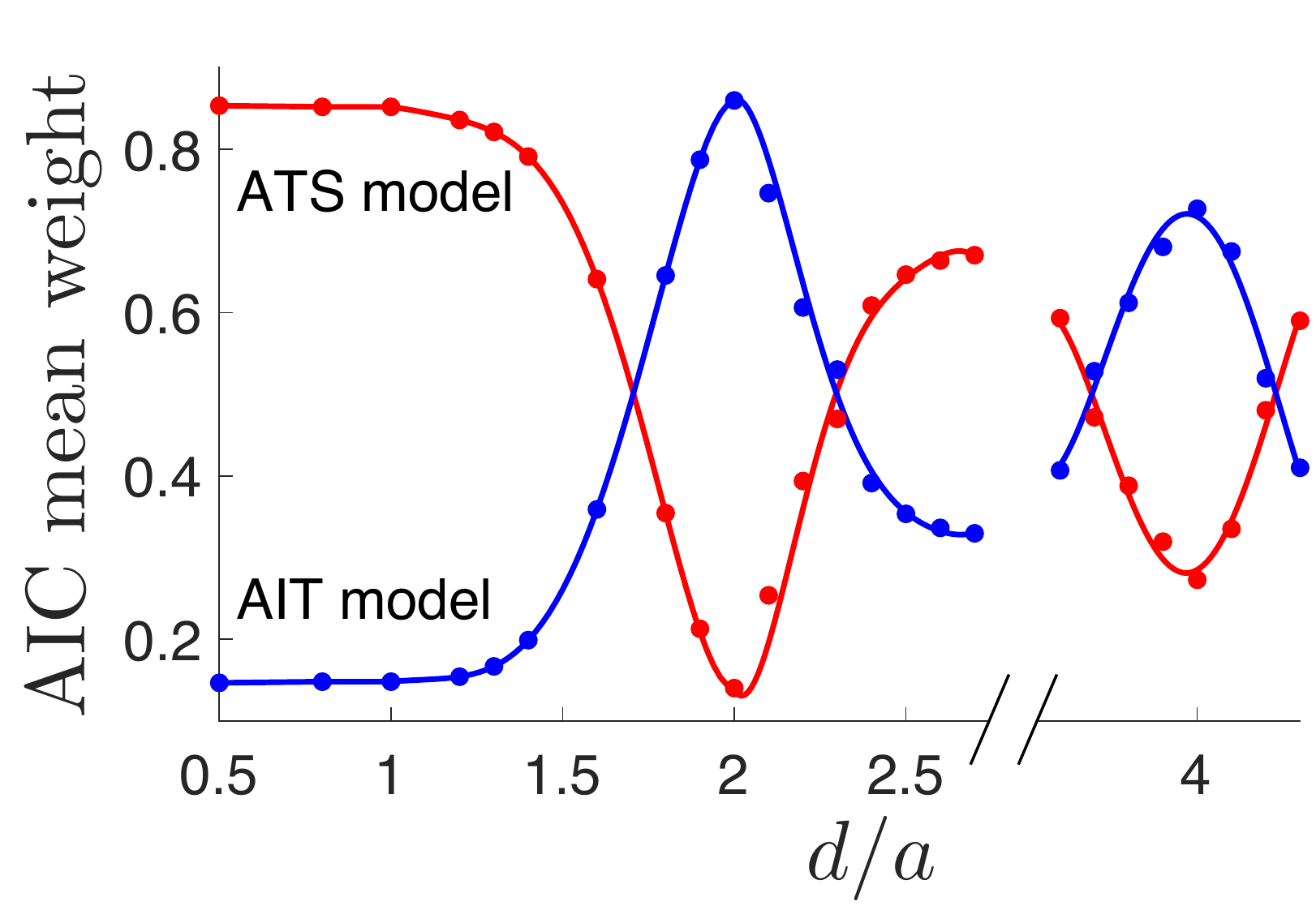}
	\caption{AIC mean weight as a function of the distance between pillars $d$ for ATS model (red line) and AIT model (blue line), in the case of $r_{2}=0.202a$. $r_{1}=0.2a$, $a=2\mu\text{m}$}
	\label{AIC}
\end{figure}

By fitting the numerical data to the model fits of $T_{ATS}$ and $T_{AIT}$ for different $r_{2}$ ($r_{1}$ fixed), the relation between the fits parameters and $r_{2}$ can be obtained as shown in Fig~\ref{fitpara}, where Fig~\ref{fitpara}(a) presents the fit parameters of the ATS model for $d=0.5a$ and Fig~\ref{fitpara}(b) are those of the AIT model for $d=2a$.
The parameters of ATS model describing each Lorentzian curve are independent of each other. It can be seen that $\Gamma_{1}$ increases with the increase of radius mismatch while $\Gamma_{2}$ increases with the increase of $r_{2}$. We think this is resulting from the symmetrical/asymmetrical vibration of the two pillars at dip1/dip2 in the ATS cases. $C_{1}$ is relatively stable while $C_{2}$ presents a slight tendency to increase.
The parameters of AIT model are related to each other. As can be seen in Fig~\ref{fitpara}(b), these four parameters all increase as the radius mismatch increases and are almost symmetrical in our range of measurement. Despite a larger values of $\Gamma_{+}$ compared with $\Gamma_{-}$, $C_{+}$ and $C_{-}$ are almost the same for different $r_{2}$ . 

With the increase of distance between pillars, the transition from ATS to AIT can be quantitatively studied by evaluating the quality of these model fits. The Akaike information criterion (AIC)\cite{burnham_model_2002} is used to discern AIT from ATS, which provides a method to select the best model from a set of models. This criterion quantifies the amount of information lost, i.e. the degree of unfitness, and is given as $I_{j}=2k-2ln(L_{j})$, where $k=4$ is the number of unknown parameters and $L_{j}$ the maximum likelihood for the considered models, i.e. $j=$ATS or AIT. Since we already found the best fit functions of $T_{ATS}$ and $T_{AIT}$, it is sufficient to calculate the likelihood of these two functions.
Then, the AIC weight $W_{j}=e^{-I_{j}/2}/\sum_{1}^{N} e^{-I_{j}/2}$ can give the relative likelihood of a candidate model. $N$ is the number of considered model. In our case, $N=2$ as only two models are involved. Since we have more than one calculated data for each model, we utilize the AIC mean per-point weight\cite{anisimov_objectively_2011} $\overline{w}_{j}=e^{-I_{j}/2n}/\sum_{1}^{N} e^{-I_{j}/2n}$ to calculate the statistically synthesized likelihood of the candidate model. $n$ is the calculated data number. The AIC mean weight can be rewritten as:
\begin{equation}
\overline{w}_{ATS}=\frac{e^{-I_{ATS}/2n}}{e^{-I_{ATS}/2n}+e^{-I_{AIT}/2n}}
\end{equation}
with $\overline{w}_{ATS}+\overline{w}_{AIT}=1$.

Fig~\ref{AIC} shows the AIC mean weight of the ATS and AIT models as function of the distance between pillars, in the case of $r_{2}=0.202a$ and $r_{1}=0.2a$. It can be seen that, as expected, the AIC mean weight of ATS model is dominant for the small distance region, which means it is preferable to use the ATS model. When $d$ increases, the AIC mean weight of AIT model starts to increase as well, and becomes dominant for $d\in$[$1.8a$, $2.2a$]. 
Since cavity mode is not presented at the peak for $d=1.8a$, and starts to interact with the dip1 for $d=2.2a$, the AIT is hence in the distance range of $d\in$[$1.9a$, $2.1a$]. When $d$ continues to increase, the AIC mean weight of ATS becomes dominant again until $d$ increases to around $4a$, i.e. the position of the second FP resonance, where we found the second AIT position
for $d\in$[$3.9a$, $4.1a$]
. The periodicity of AIT, which is a behavior that hasn't been addressed before in the existing literature, is due to the periodic apparition of the FP resonance. The AIC mean weight of the second AIT region is smaller than that of the first AIT region, since the interaction of the system decreases with the increase of distance between pillars $d$. 
Note that the ATS in our case exists only for $d<1.4a$, therefore the transition region $d\in$[$1.4a$, $1.8a$] as well as the larger distances other than the AIT regions do not represent ATS or AIT. We can see that the theoretical distinction between ATS and AIT helps the comprehension of the analytical results.
On the other hand, this criterion is useful when we cannot theoretically rule out the AIT phenomenon. This happens for example when the radius mismatch is so large that the peak is much wider with respect to the two dips. In this case, although FP resonance is still present between two dips, the transmission spectra can not be well fitted by the AIT model and therefore cannot be ascribed to an AIT.  

\begin{figure}[t]
	\centering
	\includegraphics[width=.7\linewidth]{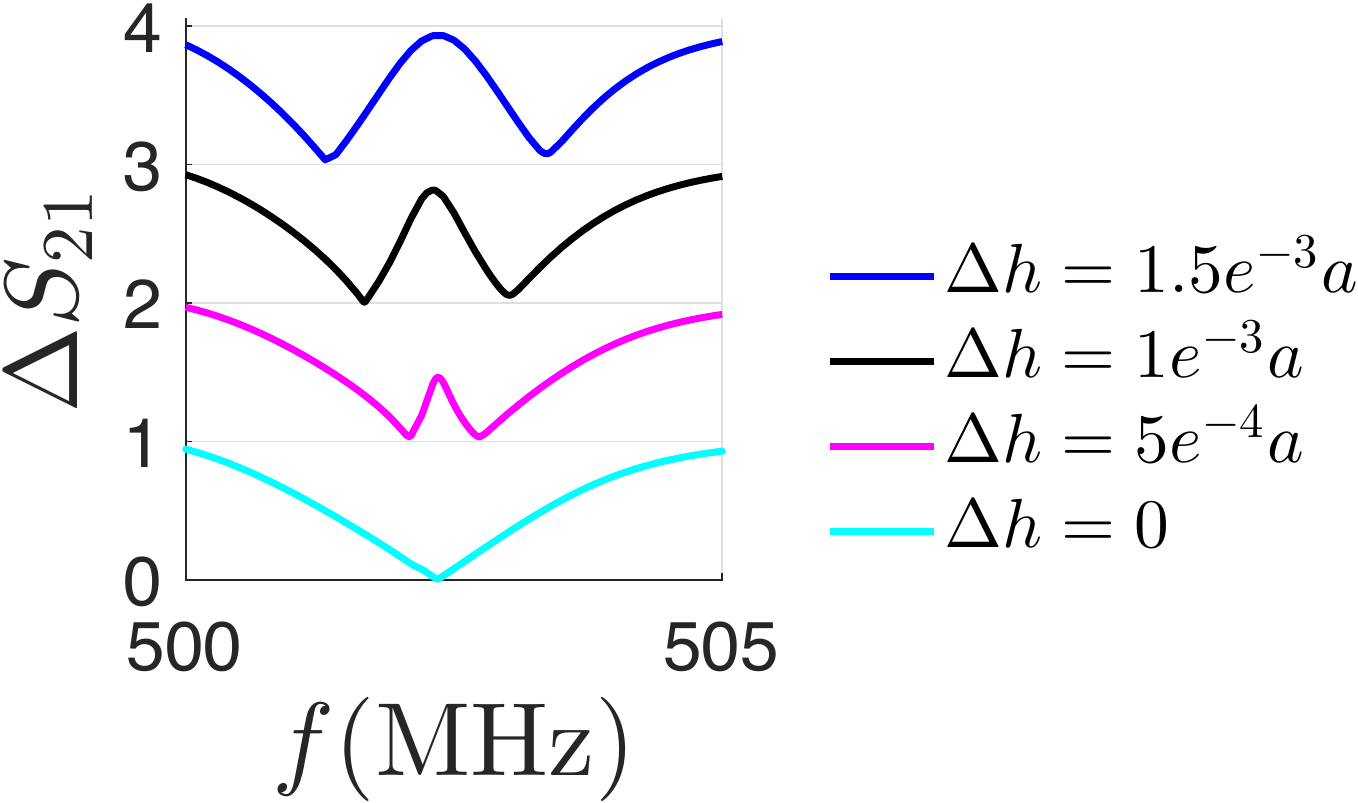}
	\caption{Normalized transmission spectra of Love waves propagating through the two lines of pillars with different heights, $h_{1}$=$h$+$\Delta h$ and $h_{2}$=$h$-$\Delta h$, in the case of $d=2a$. AIT peak rises and becomes wider with the increase of pillar height mismatch. $h=0.6a$, $r_{1,2}=0.2a$ and $a=2\mu\text{m}$}
	\label{h-AIT}
\end{figure}
The above effects can also be obtained by detuning the pillar height. The heights of the two pillars are increased/reduced from the original height $h=0.6a$ by the same amount $\Delta h$. That is, $h_{1}=h+\Delta h$ and $h_{2}=h-\Delta h$. The radius of both the two pillars $r_{1}$ and $r_{2}$ are fixed at $0.2a$. Fig~\ref{h-AIT} shows the transmission spectra of Love waves through the two lines of pillars when $d=2a$, i.e. in the region of AIT. It can be seen that by decreasing $\Delta h$, the two dips approach to each other and the peak with almost unchanged frequency decreases and becomes invisible for identical pillars. 


\section{Conclusion}
In this work, the interaction of Love waves with two lines of cylindrical Ni pillars are investigated on a silica film deposited on a 90ST quartz substrate. 
Firstly, pillar intrinsic torsional mode is demonstrated to be well excited by Love waves. One line of pillars can give rise to a sharp transmission dip due to a destructive interference. 
Secondly, acoustic analogue of Autler-Townes Splitting (ATS) and Fabry-Perot resonance of Love waves are demonstrated in two lines of identical pillars by varying the distance between the pillar lines. 
ATS appears when the distance is smaller than the half wavelength and a strong coupling is aroused between the pillar lines, causing the pillar mode induced transmission dip to split into two dips with a transparency window in the middle. This coupling decreases with the increase of pillar distance. We demonstrated the different pillar vibration symmetries at the two dip frequencies, which lead to different dip widths.
Fabry-Perot resonance exists at the positions where the distance between the pillar lines is a multiple of half wavelength. The proximity of Fabry-Perot resonance with pillar intrinsic mode gives rise to the cavity modes with transmission enhancement on the two edges of the single dip. We avoided to mix up the FP resonances with the cavity modes by presenting the different frequency variation with respect to the distance between the pillars.
Thirdly, the radius of one line of pillar is modified to detune the pillar resonant frequency. In the pillar coupling region, the coupling effect decreases with the increase of radius mismatch, and the two dips will rejoin their individual pillar mode frequencies. 
When the distance between the pillar lines is a multiple of half wavelength, Fabry-Perot resonance along with the two different pillars' resonances give rise to the Acoustically Induced Transparency (AIT). Same phenomena can also be obtained by detuning the pillar height.
Then, with similar transparency window in the transmission spectra, ATS and AIT phenomena are first fitted with the corresponding formula models, showing good agreements. The fit parameters are demonstrated as functions of the geometrical parameter. 
The Akaike information criterion (AIC) is then used to quantitatively evaluate the quality of the fit models, which illustrates the transition from ATS to AIT as well as the periodicity of AIT by increasing the distance between the pillar lines. The theoretical and analytical differentiation of ATS and AIT should be used together to discriminate the assignment of the observed spectrum to one or the other physical mechanism. The results presented in this study could be used to potential acoustic applications such as wave control, meta-materials and bio-sensors. 

\appendix
\section{Dissipation consideration}
\begin{figure}[t]
	\centering
	\includegraphics[width=\linewidth]{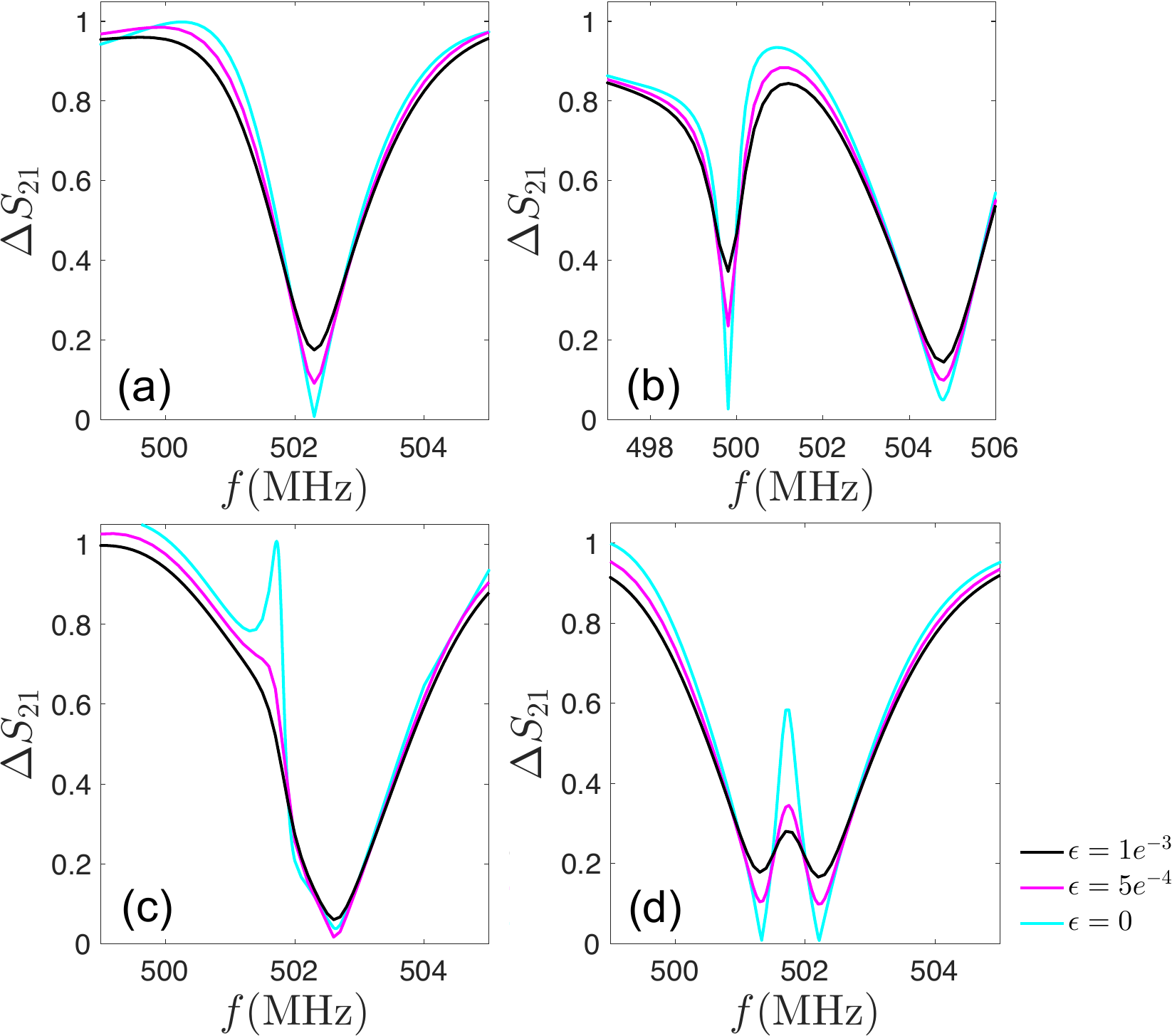}
	\caption{(a)Transmission dip for a single pillar line with $r$ =0.2$a$; (b) ATS for two lines of identical pillars with $r_{1,2}=0.2a$; (c) Cavity mode for two lines of identical pillars with $r_{1,2}=0.2a$; (d) AIT for two lines of dissimilar pillars with $r_{1}=0.2a$ and $r_{2}=0.202a$. $a=2\mu\text{m}$. Cyan curves are the transmission without dissipation. Rose and black curves correspond to the transmission with $\epsilon$= $5\times10^{-4}$ and $1\times10^{-3}$, respectively.}
	\label{dissipation}
\end{figure}

To consider the losses by dissipation, we recalculated the critical transmission curves by changing the Young's modulus $E$ of the silica film and the pillars to complex numbers $E(1+j\epsilon)$, with $\epsilon$ equals to $5\times10^{-4}$ and $1\times10^{-3}$. 
Corresponding transmission spectra for different phenomena (dips, ATS, cavity modes and AIT) are shown in Fig~\ref{dissipation}. It can be seen that all the main resonances and dips are visible for $\epsilon$ below $1\times10^{-3}$. With $\epsilon$=$1\times10^{-3}$, we can estimate a propagation length around 2mm. In experiments, Love waves can propagate for a few mm. We have already realized 500 MHz devices with distances between transmitter and receiver of a few mm\cite{talbi_simulation_2015}. This means the parameter $\epsilon$ is below $1\times10^{-3}$, which indicates that our system is robust for practical applications. 

\section{Cavity mode and FP resonance}
\subsection{two lines of identical pillars}
\begin{figure}[h!]
	\centering
	\includegraphics[width=.8\linewidth]{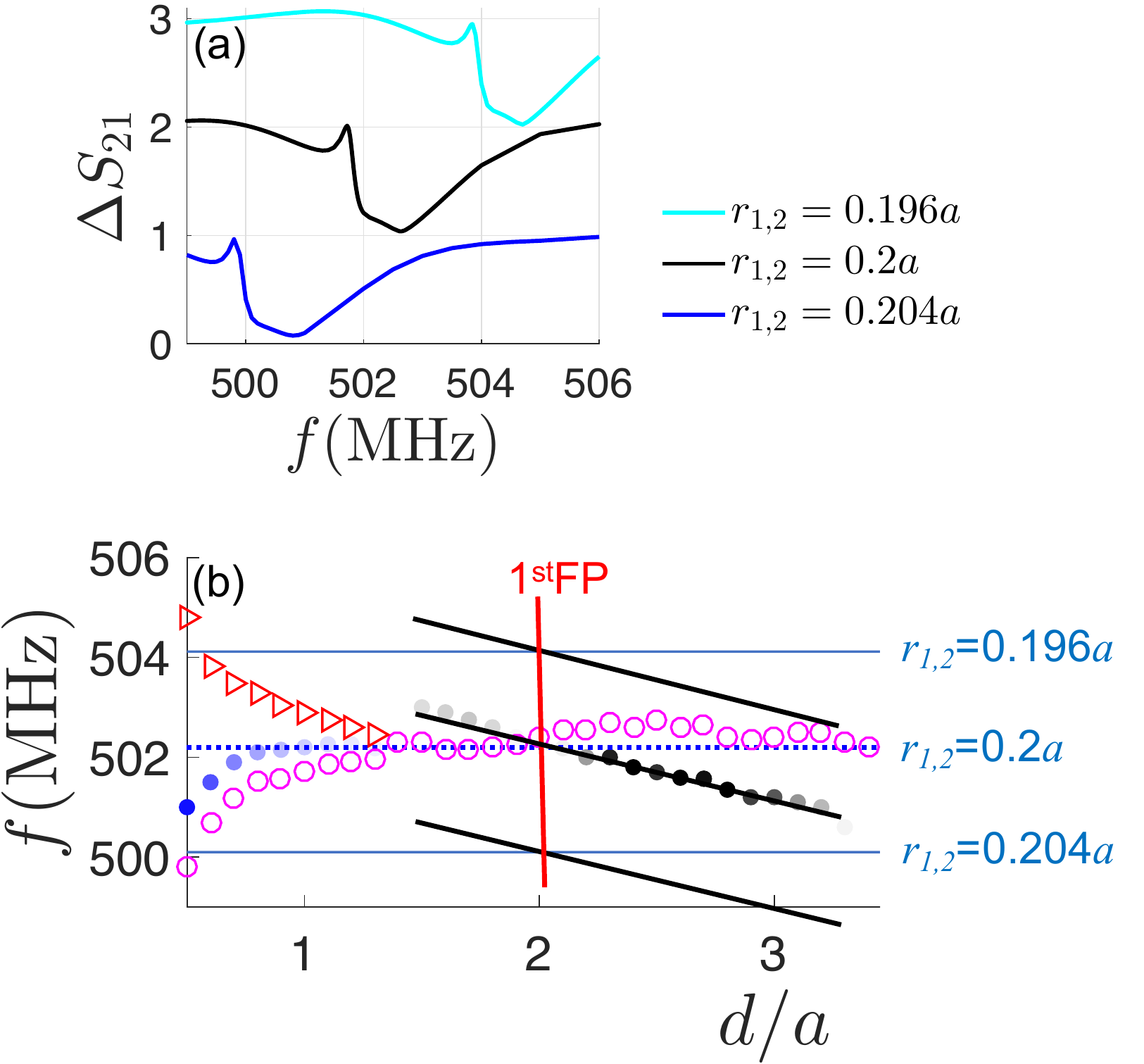}
	\caption{(a)Transmission spectra for $d=2.4a$ when all the pillars' radius vary from 0.196$a$ to 0.204$a$. (b) Cavity modes (black lines) for two lines of identical pillars when changing the pillar resonant frequency (blue lines). Red line denotes the $1^{st}$ FP resonance.}
	\label{FPCav}
\end{figure}

In order to show the behavior of the cavity modes at the vicinity of the transmission dip, we plotted in Fig~\ref{FPCav}(a) the transmission spectra for $d=2.4a$ when all the pillars' radius vary from 0.196$a$ to 0.204$a$. It can be seen that we obtain almost the same cavity mode besides the dip at different frequencies. Along with Fig~\ref{FPtrans}(b) and Fig~\ref{FPCav}(a), we can present in Fig~\ref{FPCav}(b) the relation between cavity modes (black lines) and FP resonance (red line). When we change the pillar resonant frequency (blue lines), FP resonance remains at almost the same position. For each radius value, the FP resonance is a particular case of the cavity modes when the later coincide with the dip.

\subsection{two lines of dissimilar pillars}
\begin{figure}[h!]
	\centering
	\includegraphics[width=\linewidth]{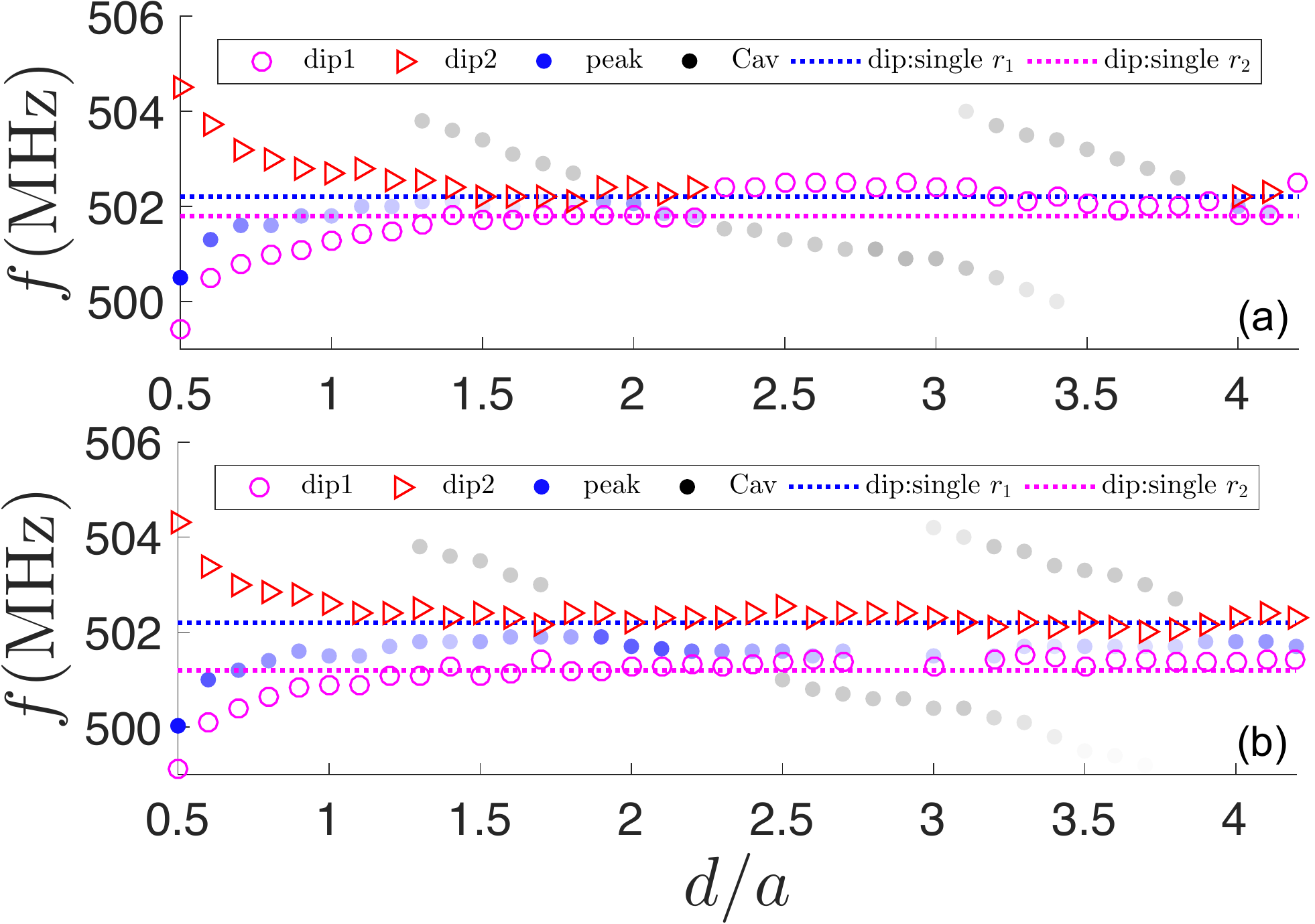}
	\caption{Frequencies of dips, peaks and cavity modes as a function of the distance $d$ between two dissimilar pillar lines, in the case of (a) $r_{1}=0.2a$ and $r_{2}=0.201a$, and (b) $r_{1}=0.2a$ and $r_{2}=0.202a$. $a=2\mu\text{m}$. }
	\label{f-d}
\end{figure}

Fig~\ref{f-d} gives the same results as in Fig~\ref{f-ec}, but with two lines of dissimilar pillars. Fig~\ref{f-d}(a) is the case of $r_{1}=0.2a$ and $r_{2}=0.201a$. Dip1 (dip2) corresponds to the second (first) pillar. It can be seen that outside the ATS region, dip2 becomes invisible after $d$ exceeding 2.2$a$. This is because on the one hand, the interaction between the two pillars becomes too weak and on the other hand, the radius mismatch is too small. This dip become visible again when $d$ approaches to $4a$, since the interaction between the two pillars regains its strength in the AIT region. When we enlarge the radius difference (see Fig~\ref{f-d}(b)), most of the dip1 in this region becomes visible. 
Moreover, we can see that the cavity modes cross the center of the two dips at $d$=$2a$ and $4a$, where we obtain the AIT resonance of FP type. This behavior is compatible with the cavity mode-FP resonance relation that we presented in Fig~\ref{FPCav}(b). FP resonances are particular cases of the cavity modes when the later fall between the two dips.

\bibliography{PnC}

\begin{thebibliography}{66}%
\makeatletter
\providecommand \@ifxundefined [1]{%
 \@ifx{#1\undefined}
}%
\providecommand \@ifnum [1]{%
 \ifnum #1\expandafter \@firstoftwo
 \else \expandafter \@secondoftwo
 \fi
}%
\providecommand \@ifx [1]{%
 \ifx #1\expandafter \@firstoftwo
 \else \expandafter \@secondoftwo
 \fi
}%
\providecommand \natexlab [1]{#1}%
\providecommand \enquote  [1]{``#1''}%
\providecommand \bibnamefont  [1]{#1}%
\providecommand \bibfnamefont [1]{#1}%
\providecommand \citenamefont [1]{#1}%
\providecommand \href@noop [0]{\@secondoftwo}%
\providecommand \href [0]{\begingroup \@sanitize@url \@href}%
\providecommand \@href[1]{\@@startlink{#1}\@@href}%
\providecommand \@@href[1]{\endgroup#1\@@endlink}%
\providecommand \@sanitize@url [0]{\catcode `\\12\catcode `\$12\catcode
  `\&12\catcode `\#12\catcode `\^12\catcode `\_12\catcode `\%12\relax}%
\providecommand \@@startlink[1]{}%
\providecommand \@@endlink[0]{}%
\providecommand \url  [0]{\begingroup\@sanitize@url \@url }%
\providecommand \@url [1]{\endgroup\@href {#1}{\urlprefix }}%
\providecommand \urlprefix  [0]{URL }%
\providecommand \Eprint [0]{\href }%
\providecommand \doibase [0]{http://dx.doi.org/}%
\providecommand \selectlanguage [0]{\@gobble}%
\providecommand \bibinfo  [0]{\@secondoftwo}%
\providecommand \bibfield  [0]{\@secondoftwo}%
\providecommand \translation [1]{[#1]}%
\providecommand \BibitemOpen [0]{}%
\providecommand \bibitemStop [0]{}%
\providecommand \bibitemNoStop [0]{.\EOS\space}%
\providecommand \EOS [0]{\spacefactor3000\relax}%
\providecommand \BibitemShut  [1]{\csname bibitem#1\endcsname}%
\let\auto@bib@innerbib\@empty
\bibitem [{\citenamefont {Fleischhauer}\ \emph {et~al.}(2005)\citenamefont
  {Fleischhauer}, \citenamefont {Imamoglu},\ and\ \citenamefont
  {Marangos}}]{fleischhauer_electromagnetically_2005}%
  \BibitemOpen
  \bibfield  {author} {\bibinfo {author} {\bibfnamefont {M.}~\bibnamefont
  {Fleischhauer}}, \bibinfo {author} {\bibfnamefont {A.}~\bibnamefont
  {Imamoglu}}, \ and\ \bibinfo {author} {\bibfnamefont {J.~P.}\ \bibnamefont
  {Marangos}},\ }\href@noop {} {\bibfield  {journal} {\bibinfo  {journal}
  {Review of Modern Physics}\ }\textbf {\bibinfo {volume} {77}},\ \bibinfo
  {pages} {633} (\bibinfo {year} {2005})}\BibitemShut {NoStop}%
\bibitem [{\citenamefont {Hau}\ \emph {et~al.}(1999)\citenamefont {Hau},
  \citenamefont {Harris}, \citenamefont {Dutton},\ and\ \citenamefont
  {Behroozi}}]{hau_light_1999}%
  \BibitemOpen
  \bibfield  {author} {\bibinfo {author} {\bibfnamefont {L.~V.}\ \bibnamefont
  {Hau}}, \bibinfo {author} {\bibfnamefont {S.~E.}\ \bibnamefont {Harris}},
  \bibinfo {author} {\bibfnamefont {Z.}~\bibnamefont {Dutton}}, \ and\ \bibinfo
  {author} {\bibfnamefont {C.~H.}\ \bibnamefont {Behroozi}},\ }\href@noop {}
  {\bibfield  {journal} {\bibinfo  {journal} {Nature}\ }\textbf {\bibinfo
  {volume} {397}},\ \bibinfo {pages} {594} (\bibinfo {year}
  {1999})}\BibitemShut {NoStop}%
\bibitem [{\citenamefont {Alotaibi}\ and\ \citenamefont
  {Sanders}(2016)}]{alotaibi_enhanced_2016}%
  \BibitemOpen
  \bibfield  {author} {\bibinfo {author} {\bibfnamefont {H.~M.~M.}\
  \bibnamefont {Alotaibi}}\ and\ \bibinfo {author} {\bibfnamefont {B.~C.}\
  \bibnamefont {Sanders}},\ }\href@noop {} {\bibfield  {journal} {\bibinfo
  {journal} {Physical Review A}\ }\textbf {\bibinfo {volume} {94}},\ \bibinfo
  {pages} {053832} (\bibinfo {year} {2016})}\BibitemShut {NoStop}%
\bibitem [{\citenamefont {Heinze}\ \emph {et~al.}(2013)\citenamefont {Heinze},
  \citenamefont {Hubrich},\ and\ \citenamefont
  {Halfmann}}]{heinze_stopped_2013}%
  \BibitemOpen
  \bibfield  {author} {\bibinfo {author} {\bibfnamefont {G.}~\bibnamefont
  {Heinze}}, \bibinfo {author} {\bibfnamefont {C.}~\bibnamefont {Hubrich}}, \
  and\ \bibinfo {author} {\bibfnamefont {T.}~\bibnamefont {Halfmann}},\
  }\href@noop {} {\bibfield  {journal} {\bibinfo  {journal} {Physical Review
  Letters}\ }\textbf {\bibinfo {volume} {111}},\ \bibinfo {pages} {033601}
  (\bibinfo {year} {2013})}\BibitemShut {NoStop}%
\bibitem [{\citenamefont {Autler}\ and\ \citenamefont
  {Townes}(1955)}]{autler_stark_1955}%
  \BibitemOpen
  \bibfield  {author} {\bibinfo {author} {\bibfnamefont {S.~H.}\ \bibnamefont
  {Autler}}\ and\ \bibinfo {author} {\bibfnamefont {C.~H.}\ \bibnamefont
  {Townes}},\ }\href@noop {} {\bibfield  {journal} {\bibinfo  {journal}
  {Physical Review Journal Archive}\ }\textbf {\bibinfo {volume} {100}},\
  \bibinfo {pages} {703} (\bibinfo {year} {1955})}\BibitemShut {NoStop}%
\bibitem [{\citenamefont
  {Abi-Salloum}(2010)}]{abisalloum_electromagnetically_2010}%
  \BibitemOpen
  \bibfield  {author} {\bibinfo {author} {\bibfnamefont {T.~Y.}\ \bibnamefont
  {Abi-Salloum}},\ }\href@noop {} {\bibfield  {journal} {\bibinfo  {journal}
  {Physical Review A}\ }\textbf {\bibinfo {volume} {81}},\ \bibinfo {pages}
  {053836} (\bibinfo {year} {2010})}\BibitemShut {NoStop}%
\bibitem [{\citenamefont {Boller}\ \emph {et~al.}(1991)\citenamefont {Boller},
  \citenamefont {Imamoglu},\ and\ \citenamefont
  {Harris}}]{boller_observation_1991}%
  \BibitemOpen
  \bibfield  {author} {\bibinfo {author} {\bibfnamefont {K.~J.}\ \bibnamefont
  {Boller}}, \bibinfo {author} {\bibfnamefont {A.}~\bibnamefont {Imamoglu}}, \
  and\ \bibinfo {author} {\bibfnamefont {S.~E.}\ \bibnamefont {Harris}},\
  }\href@noop {} {\bibfield  {journal} {\bibinfo  {journal} {Physical Review
  Letters}\ }\textbf {\bibinfo {volume} {66}},\ \bibinfo {pages} {2593}
  (\bibinfo {year} {1991})}\BibitemShut {NoStop}%
\bibitem [{\citenamefont {Peng}\ \emph {et~al.}(2014)\citenamefont {Peng},
  \citenamefont {Ozdemir}, \citenamefont {Chen}, \citenamefont {Nori},\ and\
  \citenamefont {Yang}}]{peng_what_2014}%
  \BibitemOpen
  \bibfield  {author} {\bibinfo {author} {\bibfnamefont {B.}~\bibnamefont
  {Peng}}, \bibinfo {author} {\bibfnamefont {S.~K.}\ \bibnamefont {Ozdemir}},
  \bibinfo {author} {\bibfnamefont {W.}~\bibnamefont {Chen}}, \bibinfo {author}
  {\bibfnamefont {F.}~\bibnamefont {Nori}}, \ and\ \bibinfo {author}
  {\bibfnamefont {L.}~\bibnamefont {Yang}},\ }\href@noop {} {\bibfield
  {journal} {\bibinfo  {journal} {Nature Communications}\ }\textbf {\bibinfo
  {volume} {5}},\ \bibinfo {pages} {5082} (\bibinfo {year} {2014})}\BibitemShut
  {NoStop}%
\bibitem [{\citenamefont {Liu}\ \emph {et~al.}(2017)\citenamefont {Liu},
  \citenamefont {Li},\ and\ \citenamefont
  {Xiao}}]{liu_electromagnetically_2017}%
  \BibitemOpen
  \bibfield  {author} {\bibinfo {author} {\bibfnamefont {Y.-C.}\ \bibnamefont
  {Liu}}, \bibinfo {author} {\bibfnamefont {B.-B.}\ \bibnamefont {Li}}, \ and\
  \bibinfo {author} {\bibfnamefont {Y.-F.}\ \bibnamefont {Xiao}},\ }\href@noop
  {} {\bibfield  {journal} {\bibinfo  {journal} {Nanophotonics}\ }\textbf
  {\bibinfo {volume} {6}},\ \bibinfo {pages} {789} (\bibinfo {year}
  {2017})}\BibitemShut {NoStop}%
\bibitem [{\citenamefont {Wei}\ and\ \citenamefont
  {Jian}(2017)}]{wei_objectively_2017}%
  \BibitemOpen
  \bibfield  {author} {\bibinfo {author} {\bibfnamefont {B.}~\bibnamefont
  {Wei}}\ and\ \bibinfo {author} {\bibfnamefont {S.}~\bibnamefont {Jian}},\
  }\href@noop {} {\bibfield  {journal} {\bibinfo  {journal} {Journal of
  optics}\ }\textbf {\bibinfo {volume} {19}},\ \bibinfo {pages} {115001}
  (\bibinfo {year} {2017})}\BibitemShut {NoStop}%
\bibitem [{\citenamefont {Dong}\ \emph {et~al.}(2012)\citenamefont {Dong},
  \citenamefont {Fiore}, \citenamefont {Kuzyk},\ and\ \citenamefont
  {Wang}}]{dong_optomechanical_2012}%
  \BibitemOpen
  \bibfield  {author} {\bibinfo {author} {\bibfnamefont {C.}~\bibnamefont
  {Dong}}, \bibinfo {author} {\bibfnamefont {V.}~\bibnamefont {Fiore}},
  \bibinfo {author} {\bibfnamefont {M.~C.}\ \bibnamefont {Kuzyk}}, \ and\
  \bibinfo {author} {\bibfnamefont {H.}~\bibnamefont {Wang}},\ }\href@noop {}
  {\bibfield  {journal} {\bibinfo  {journal} {Science}\ }\textbf {\bibinfo
  {volume} {21}},\ \bibinfo {pages} {1609} (\bibinfo {year}
  {2012})}\BibitemShut {NoStop}%
\bibitem [{\citenamefont {Safavi-Naeini}\ \emph {et~al.}(2011)\citenamefont
  {Safavi-Naeini}, \citenamefont {Mayer~Alegre}, \citenamefont {Chan},
  \citenamefont {Eichenfield}, \citenamefont {Winger}, \citenamefont {Lin},
  \citenamefont {Hill}, \citenamefont {Chang},\ and\ \citenamefont
  {Painter}}]{safavi_electromagnetically_2011}%
  \BibitemOpen
  \bibfield  {author} {\bibinfo {author} {\bibfnamefont {A.~H.}\ \bibnamefont
  {Safavi-Naeini}}, \bibinfo {author} {\bibfnamefont {T.~P.}\ \bibnamefont
  {Mayer~Alegre}}, \bibinfo {author} {\bibfnamefont {J.}~\bibnamefont {Chan}},
  \bibinfo {author} {\bibfnamefont {M.}~\bibnamefont {Eichenfield}}, \bibinfo
  {author} {\bibfnamefont {M.}~\bibnamefont {Winger}}, \bibinfo {author}
  {\bibfnamefont {Q.}~\bibnamefont {Lin}}, \bibinfo {author} {\bibfnamefont
  {J.~T.}\ \bibnamefont {Hill}}, \bibinfo {author} {\bibfnamefont {D.~E.}\
  \bibnamefont {Chang}}, \ and\ \bibinfo {author} {\bibfnamefont
  {O.}~\bibnamefont {Painter}},\ }\href@noop {} {\bibfield  {journal} {\bibinfo
   {journal} {Nature}\ }\textbf {\bibinfo {volume} {472}},\ \bibinfo {pages}
  {69} (\bibinfo {year} {2011})}\BibitemShut {NoStop}%
\bibitem [{\citenamefont {Weis}\ \emph {et~al.}(2010)\citenamefont {Weis},
  \citenamefont {Riviere}, \citenamefont {Deleglise}, \citenamefont {Gavartin},
  \citenamefont {Arcizet}, \citenamefont {Schliesser},\ and\ \citenamefont
  {Kippenberg}}]{weis_optomechanically_2010}%
  \BibitemOpen
  \bibfield  {author} {\bibinfo {author} {\bibfnamefont {S.}~\bibnamefont
  {Weis}}, \bibinfo {author} {\bibfnamefont {R.}~\bibnamefont {Riviere}},
  \bibinfo {author} {\bibfnamefont {S.}~\bibnamefont {Deleglise}}, \bibinfo
  {author} {\bibfnamefont {E.}~\bibnamefont {Gavartin}}, \bibinfo {author}
  {\bibfnamefont {O.}~\bibnamefont {Arcizet}}, \bibinfo {author} {\bibfnamefont
  {A.}~\bibnamefont {Schliesser}}, \ and\ \bibinfo {author} {\bibfnamefont
  {T.~J.}\ \bibnamefont {Kippenberg}},\ }\href@noop {} {\bibfield  {journal}
  {\bibinfo  {journal} {Science}\ }\textbf {\bibinfo {volume} {330}},\ \bibinfo
  {pages} {1520} (\bibinfo {year} {2010})}\BibitemShut {NoStop}%
\bibitem [{\citenamefont {Zhang}\ \emph {et~al.}(2008)\citenamefont {Zhang},
  \citenamefont {Genov}, \citenamefont {Wang}, \citenamefont {Liu},\ and\
  \citenamefont {Zhang}}]{zhang_plasmon_2008}%
  \BibitemOpen
  \bibfield  {author} {\bibinfo {author} {\bibfnamefont {S.}~\bibnamefont
  {Zhang}}, \bibinfo {author} {\bibfnamefont {D.~A.}\ \bibnamefont {Genov}},
  \bibinfo {author} {\bibfnamefont {Y.}~\bibnamefont {Wang}}, \bibinfo {author}
  {\bibfnamefont {M.}~\bibnamefont {Liu}}, \ and\ \bibinfo {author}
  {\bibfnamefont {X.}~\bibnamefont {Zhang}},\ }\href@noop {} {\bibfield
  {journal} {\bibinfo  {journal} {Physical Review Letters}\ }\textbf {\bibinfo
  {volume} {101}},\ \bibinfo {pages} {047401} (\bibinfo {year}
  {2008})}\BibitemShut {NoStop}%
\bibitem [{\citenamefont {Liu}\ \emph {et~al.}(2009)\citenamefont {Liu},
  \citenamefont {Langguth}, \citenamefont {Weiss}, \citenamefont {Kastel},
  \citenamefont {Fleischhauer}, \citenamefont {Pfau},\ and\ \citenamefont
  {Giessen}}]{liu_plasmonic_2009}%
  \BibitemOpen
  \bibfield  {author} {\bibinfo {author} {\bibfnamefont {N.}~\bibnamefont
  {Liu}}, \bibinfo {author} {\bibfnamefont {L.}~\bibnamefont {Langguth}},
  \bibinfo {author} {\bibfnamefont {T.}~\bibnamefont {Weiss}}, \bibinfo
  {author} {\bibfnamefont {J.}~\bibnamefont {Kastel}}, \bibinfo {author}
  {\bibfnamefont {M.}~\bibnamefont {Fleischhauer}}, \bibinfo {author}
  {\bibfnamefont {T.}~\bibnamefont {Pfau}}, \ and\ \bibinfo {author}
  {\bibfnamefont {H.}~\bibnamefont {Giessen}},\ }\href@noop {} {\bibfield
  {journal} {\bibinfo  {journal} {Nature Materials}\ }\textbf {\bibinfo
  {volume} {8}},\ \bibinfo {pages} {758} (\bibinfo {year} {2009})}\BibitemShut
  {NoStop}%
\bibitem [{\citenamefont {Han}\ and\ \citenamefont
  {Bozhevolnyi}(2011)}]{Han_plasmon_2011}%
  \BibitemOpen
  \bibfield  {author} {\bibinfo {author} {\bibfnamefont {Z.}~\bibnamefont
  {Han}}\ and\ \bibinfo {author} {\bibfnamefont {S.~I.}\ \bibnamefont
  {Bozhevolnyi}},\ }\href@noop {} {\bibfield  {journal} {\bibinfo  {journal}
  {Opt. Express}\ }\textbf {\bibinfo {volume} {19}},\ \bibinfo {pages} {3251}
  (\bibinfo {year} {2011})}\BibitemShut {NoStop}%
\bibitem [{\citenamefont {Liu}\ \emph {et~al.}(2010{\natexlab{a}})\citenamefont
  {Liu}, \citenamefont {Weiss}, \citenamefont {Mesch}, \citenamefont
  {Langguth}, \citenamefont {Eigenthaler}, \citenamefont {Hirscher},
  \citenamefont {Sonnichsen},\ and\ \citenamefont {Giessen}}]{liu_planar_2010}%
  \BibitemOpen
  \bibfield  {author} {\bibinfo {author} {\bibfnamefont {N.}~\bibnamefont
  {Liu}}, \bibinfo {author} {\bibfnamefont {T.}~\bibnamefont {Weiss}}, \bibinfo
  {author} {\bibfnamefont {M.}~\bibnamefont {Mesch}}, \bibinfo {author}
  {\bibfnamefont {L.}~\bibnamefont {Langguth}}, \bibinfo {author}
  {\bibfnamefont {U.}~\bibnamefont {Eigenthaler}}, \bibinfo {author}
  {\bibfnamefont {M.}~\bibnamefont {Hirscher}}, \bibinfo {author}
  {\bibfnamefont {C.}~\bibnamefont {Sonnichsen}}, \ and\ \bibinfo {author}
  {\bibfnamefont {H.}~\bibnamefont {Giessen}},\ }\href@noop {} {\bibfield
  {journal} {\bibinfo  {journal} {Nano Letters}\ }\textbf {\bibinfo {volume}
  {10}},\ \bibinfo {pages} {1103} (\bibinfo {year}
  {2010}{\natexlab{a}})}\BibitemShut {NoStop}%
\bibitem [{\citenamefont {Gu}\ \emph {et~al.}(2012)\citenamefont {Gu},
  \citenamefont {Singh}, \citenamefont {Liu}, \citenamefont {Zhang},
  \citenamefont {Ma}, \citenamefont {Zhang}, \citenamefont {Maier},
  \citenamefont {Tian}, \citenamefont {Azad}, \citenamefont {Chen},
  \citenamefont {Taylor}, \citenamefont {Han},\ and\ \citenamefont
  {Zhang}}]{gu_active_2012}%
  \BibitemOpen
  \bibfield  {author} {\bibinfo {author} {\bibfnamefont {J.}~\bibnamefont
  {Gu}}, \bibinfo {author} {\bibfnamefont {R.}~\bibnamefont {Singh}}, \bibinfo
  {author} {\bibfnamefont {X.}~\bibnamefont {Liu}}, \bibinfo {author}
  {\bibfnamefont {X.}~\bibnamefont {Zhang}}, \bibinfo {author} {\bibfnamefont
  {Y.}~\bibnamefont {Ma}}, \bibinfo {author} {\bibfnamefont {S.}~\bibnamefont
  {Zhang}}, \bibinfo {author} {\bibfnamefont {S.~A.}\ \bibnamefont {Maier}},
  \bibinfo {author} {\bibfnamefont {Z.}~\bibnamefont {Tian}}, \bibinfo {author}
  {\bibfnamefont {A.~K.}\ \bibnamefont {Azad}}, \bibinfo {author}
  {\bibfnamefont {H.-T.}\ \bibnamefont {Chen}}, \bibinfo {author}
  {\bibfnamefont {A.~J.}\ \bibnamefont {Taylor}}, \bibinfo {author}
  {\bibfnamefont {J.}~\bibnamefont {Han}}, \ and\ \bibinfo {author}
  {\bibfnamefont {W.}~\bibnamefont {Zhang}},\ }\href@noop {} {\bibfield
  {journal} {\bibinfo  {journal} {Nature Communications}\ }\textbf {\bibinfo
  {volume} {3}},\ \bibinfo {pages} {1151} (\bibinfo {year} {2012})}\BibitemShut
  {NoStop}%
\bibitem [{\citenamefont {Papasimakis}\ \emph {et~al.}(2008)\citenamefont
  {Papasimakis}, \citenamefont {Fedotov}, \citenamefont {Zheludev},\ and\
  \citenamefont {Prosvirnin}}]{papasimakis_metamaterial_2008}%
  \BibitemOpen
  \bibfield  {author} {\bibinfo {author} {\bibfnamefont {N.}~\bibnamefont
  {Papasimakis}}, \bibinfo {author} {\bibfnamefont {V.~A.}\ \bibnamefont
  {Fedotov}}, \bibinfo {author} {\bibfnamefont {N.~I.}\ \bibnamefont
  {Zheludev}}, \ and\ \bibinfo {author} {\bibfnamefont {S.~L.}\ \bibnamefont
  {Prosvirnin}},\ }\href@noop {} {\bibfield  {journal} {\bibinfo  {journal}
  {Physical Review Letters}\ }\textbf {\bibinfo {volume} {101}},\ \bibinfo
  {pages} {253903} (\bibinfo {year} {2008})}\BibitemShut {NoStop}%
\bibitem [{\citenamefont {Anisimov}\ \emph {et~al.}(2011)\citenamefont
  {Anisimov}, \citenamefont {Dowling},\ and\ \citenamefont
  {Sanders}}]{anisimov_objectively_2011}%
  \BibitemOpen
  \bibfield  {author} {\bibinfo {author} {\bibfnamefont {P.~M.}\ \bibnamefont
  {Anisimov}}, \bibinfo {author} {\bibfnamefont {J.~P.}\ \bibnamefont
  {Dowling}}, \ and\ \bibinfo {author} {\bibfnamefont {B.~C.}\ \bibnamefont
  {Sanders}},\ }\href@noop {} {\bibfield  {journal} {\bibinfo  {journal}
  {Physical Review Letters}\ }\textbf {\bibinfo {volume} {107}},\ \bibinfo
  {pages} {163604} (\bibinfo {year} {2011})}\BibitemShut {NoStop}%
\bibitem [{\citenamefont {Sun}\ \emph {et~al.}(2014)\citenamefont {Sun},
  \citenamefont {Liu}, \citenamefont {Ian}, \citenamefont {You}, \citenamefont
  {Il'ichev},\ and\ \citenamefont {Nori}}]{sun_electromagnetically_2014}%
  \BibitemOpen
  \bibfield  {author} {\bibinfo {author} {\bibfnamefont {H.-C.}\ \bibnamefont
  {Sun}}, \bibinfo {author} {\bibfnamefont {Y.-X.}\ \bibnamefont {Liu}},
  \bibinfo {author} {\bibfnamefont {H.}~\bibnamefont {Ian}}, \bibinfo {author}
  {\bibfnamefont {J.~Q.}\ \bibnamefont {You}}, \bibinfo {author} {\bibfnamefont
  {E.}~\bibnamefont {Il'ichev}}, \ and\ \bibinfo {author} {\bibfnamefont
  {F.}~\bibnamefont {Nori}},\ }\href@noop {} {\bibfield  {journal} {\bibinfo
  {journal} {Physical Review A}\ }\textbf {\bibinfo {volume} {89}},\ \bibinfo
  {pages} {063822} (\bibinfo {year} {2014})}\BibitemShut {NoStop}%
\bibitem [{\citenamefont {Liu}\ \emph {et~al.}(2016)\citenamefont {Liu},
  \citenamefont {Yang}, \citenamefont {Wang}, \citenamefont {Xu},\ and\
  \citenamefont {Xiao}}]{liu_experimental_2016}%
  \BibitemOpen
  \bibfield  {author} {\bibinfo {author} {\bibfnamefont {J.}~\bibnamefont
  {Liu}}, \bibinfo {author} {\bibfnamefont {H.}~\bibnamefont {Yang}}, \bibinfo
  {author} {\bibfnamefont {C.}~\bibnamefont {Wang}}, \bibinfo {author}
  {\bibfnamefont {K.}~\bibnamefont {Xu}}, \ and\ \bibinfo {author}
  {\bibfnamefont {J.}~\bibnamefont {Xiao}},\ }\href@noop {} {\bibfield
  {journal} {\bibinfo  {journal} {Scientific Reports}\ }\textbf {\bibinfo
  {volume} {6}},\ \bibinfo {pages} {19040} (\bibinfo {year}
  {2016})}\BibitemShut {NoStop}%
\bibitem [{\citenamefont {Giner}\ \emph {et~al.}(2013)\citenamefont {Giner},
  \citenamefont {Veissier}, \citenamefont {Sparkes}, \citenamefont {Sheremet},
  \citenamefont {Nicolas}, \citenamefont {Mishina}, \citenamefont {Scherman},
  \citenamefont {Burks}, \citenamefont {Shomroni}, \citenamefont {Kupriyanov},
  \citenamefont {Lam}, \citenamefont {E.},\ and\ \citenamefont
  {Laurat}}]{giner_experimental_2013}%
  \BibitemOpen
  \bibfield  {author} {\bibinfo {author} {\bibfnamefont {L.}~\bibnamefont
  {Giner}}, \bibinfo {author} {\bibfnamefont {L.}~\bibnamefont {Veissier}},
  \bibinfo {author} {\bibfnamefont {B.}~\bibnamefont {Sparkes}}, \bibinfo
  {author} {\bibfnamefont {A.}~\bibnamefont {Sheremet}}, \bibinfo {author}
  {\bibfnamefont {A.}~\bibnamefont {Nicolas}}, \bibinfo {author} {\bibfnamefont
  {O.}~\bibnamefont {Mishina}}, \bibinfo {author} {\bibfnamefont
  {M.}~\bibnamefont {Scherman}}, \bibinfo {author} {\bibfnamefont
  {S.}~\bibnamefont {Burks}}, \bibinfo {author} {\bibfnamefont
  {I.}~\bibnamefont {Shomroni}}, \bibinfo {author} {\bibfnamefont {D.~V.}\
  \bibnamefont {Kupriyanov}}, \bibinfo {author} {\bibfnamefont
  {P.}~\bibnamefont {Lam}}, \bibinfo {author} {\bibfnamefont {G.}~\bibnamefont
  {E.}}, \ and\ \bibinfo {author} {\bibfnamefont {J.}~\bibnamefont {Laurat}},\
  }\href@noop {} {\bibfield  {journal} {\bibinfo  {journal} {American Physical
  Society}\ }\textbf {\bibinfo {volume} {87}},\ \bibinfo {pages} {013823}
  (\bibinfo {year} {2013})}\BibitemShut {NoStop}%
\bibitem [{\citenamefont {Zhu}\ \emph {et~al.}(2013)\citenamefont {Zhu},
  \citenamefont {Tan},\ and\ \citenamefont {Huang}}]{zhu_crossover_2013}%
  \BibitemOpen
  \bibfield  {author} {\bibinfo {author} {\bibfnamefont {C.}~\bibnamefont
  {Zhu}}, \bibinfo {author} {\bibfnamefont {C.}~\bibnamefont {Tan}}, \ and\
  \bibinfo {author} {\bibfnamefont {G.}~\bibnamefont {Huang}},\ }\href@noop {}
  {\bibfield  {journal} {\bibinfo  {journal} {Physical Review A}\ }\textbf
  {\bibinfo {volume} {87}},\ \bibinfo {pages} {043813} (\bibinfo {year}
  {2013})}\BibitemShut {NoStop}%
\bibitem [{\citenamefont {Tan}\ and\ \citenamefont
  {Huang}(2014)}]{tan_crossover_2014}%
  \BibitemOpen
  \bibfield  {author} {\bibinfo {author} {\bibfnamefont {C.}~\bibnamefont
  {Tan}}\ and\ \bibinfo {author} {\bibfnamefont {G.}~\bibnamefont {Huang}},\
  }\href@noop {} {\bibfield  {journal} {\bibinfo  {journal} {Journal of the
  Optical Society of America B}\ }\textbf {\bibinfo {volume} {31}},\ \bibinfo
  {pages} {704} (\bibinfo {year} {2014})}\BibitemShut {NoStop}%
\bibitem [{\citenamefont {Radeonychev}\ \emph {et~al.}(2006)\citenamefont
  {Radeonychev}, \citenamefont {Tokman}, \citenamefont {Litvak},\ and\
  \citenamefont {Kocharovskaya}}]{radeonychev_acoustically_2006}%
  \BibitemOpen
  \bibfield  {author} {\bibinfo {author} {\bibfnamefont {Y.~V.}\ \bibnamefont
  {Radeonychev}}, \bibinfo {author} {\bibfnamefont {M.~D.}\ \bibnamefont
  {Tokman}}, \bibinfo {author} {\bibfnamefont {A.~G.}\ \bibnamefont {Litvak}},
  \ and\ \bibinfo {author} {\bibfnamefont {O.}~\bibnamefont {Kocharovskaya}},\
  }\href@noop {} {\bibfield  {journal} {\bibinfo  {journal} {Physical Review
  Letters}\ }\textbf {\bibinfo {volume} {96}},\ \bibinfo {pages} {093602}
  (\bibinfo {year} {2006})}\BibitemShut {NoStop}%
\bibitem [{\citenamefont {Liu}\ \emph {et~al.}(2010{\natexlab{b}})\citenamefont
  {Liu}, \citenamefont {Ke}, \citenamefont {Zhang}, \citenamefont {Wen},
  \citenamefont {Shi}, \citenamefont {Liu},\ and\ \citenamefont
  {Sheng}}]{liu_acoustic_2010}%
  \BibitemOpen
  \bibfield  {author} {\bibinfo {author} {\bibfnamefont {F.}~\bibnamefont
  {Liu}}, \bibinfo {author} {\bibfnamefont {M.}~\bibnamefont {Ke}}, \bibinfo
  {author} {\bibfnamefont {A.}~\bibnamefont {Zhang}}, \bibinfo {author}
  {\bibfnamefont {W.}~\bibnamefont {Wen}}, \bibinfo {author} {\bibfnamefont
  {J.}~\bibnamefont {Shi}}, \bibinfo {author} {\bibfnamefont {Z.}~\bibnamefont
  {Liu}}, \ and\ \bibinfo {author} {\bibfnamefont {P.}~\bibnamefont {Sheng}},\
  }\href@noop {} {\bibfield  {journal} {\bibinfo  {journal} {Physical Review
  E}\ }\textbf {\bibinfo {volume} {82}},\ \bibinfo {pages} {026601} (\bibinfo
  {year} {2010}{\natexlab{b}})}\BibitemShut {NoStop}%
\bibitem [{\citenamefont {Santillan}\ and\ \citenamefont
  {Bozhevolnyi}(2011)}]{santillan_acoustic_2011}%
  \BibitemOpen
  \bibfield  {author} {\bibinfo {author} {\bibfnamefont {A.}~\bibnamefont
  {Santillan}}\ and\ \bibinfo {author} {\bibfnamefont {S.~I.}\ \bibnamefont
  {Bozhevolnyi}},\ }\href@noop {} {\bibfield  {journal} {\bibinfo  {journal}
  {Physical Review B}\ }\textbf {\bibinfo {volume} {84}},\ \bibinfo {pages}
  {064304} (\bibinfo {year} {2011})}\BibitemShut {NoStop}%
\bibitem [{\citenamefont {Amin}\ \emph {et~al.}(2015)\citenamefont {Amin},
  \citenamefont {Elayouch}, \citenamefont {Farhat}, \citenamefont {Addouche},
  \citenamefont {Khelif},\ and\ \citenamefont
  {Bagci}}]{amin_acoustically_2015}%
  \BibitemOpen
  \bibfield  {author} {\bibinfo {author} {\bibfnamefont {M.}~\bibnamefont
  {Amin}}, \bibinfo {author} {\bibfnamefont {A.}~\bibnamefont {Elayouch}},
  \bibinfo {author} {\bibfnamefont {M.}~\bibnamefont {Farhat}}, \bibinfo
  {author} {\bibfnamefont {M.}~\bibnamefont {Addouche}}, \bibinfo {author}
  {\bibfnamefont {A.}~\bibnamefont {Khelif}}, \ and\ \bibinfo {author}
  {\bibfnamefont {H.}~\bibnamefont {Bagci}},\ }\href@noop {} {\bibfield
  {journal} {\bibinfo  {journal} {Journal of Applied Physics}\ }\textbf
  {\bibinfo {volume} {118}},\ \bibinfo {pages} {093602} (\bibinfo {year}
  {2015})}\BibitemShut {NoStop}%
\bibitem [{\citenamefont {Quotane}\ \emph {et~al.}(2018)\citenamefont
  {Quotane}, \citenamefont {El~Boudouti},\ and\ \citenamefont
  {Djafari-Rouhani}}]{quotane_trapped_2018}%
  \BibitemOpen
  \bibfield  {author} {\bibinfo {author} {\bibfnamefont {I.}~\bibnamefont
  {Quotane}}, \bibinfo {author} {\bibfnamefont {E.~H.}\ \bibnamefont
  {El~Boudouti}}, \ and\ \bibinfo {author} {\bibfnamefont {B.}~\bibnamefont
  {Djafari-Rouhani}},\ }\href@noop {} {\bibfield  {journal} {\bibinfo
  {journal} {Physical Review B}\ }\textbf {\bibinfo {volume} {97}},\ \bibinfo
  {pages} {024304} (\bibinfo {year} {2018})}\BibitemShut {NoStop}%
\bibitem [{\citenamefont {Jin}\ \emph {et~al.}(2018)\citenamefont {Jin},
  \citenamefont {Pennec},\ and\ \citenamefont
  {Djafari-Rouhani}}]{jin_acoustic_2018}%
  \BibitemOpen
  \bibfield  {author} {\bibinfo {author} {\bibfnamefont {Y.}~\bibnamefont
  {Jin}}, \bibinfo {author} {\bibfnamefont {Y.}~\bibnamefont {Pennec}}, \ and\
  \bibinfo {author} {\bibfnamefont {B.}~\bibnamefont {Djafari-Rouhani}},\
  }\href@noop {} {\bibfield  {journal} {\bibinfo  {journal} {Journal of
  Physical D: Applied Physics}\ }\textbf {\bibinfo {volume} {51}},\ \bibinfo
  {pages} {494004} (\bibinfo {year} {2018})}\BibitemShut {NoStop}%
\bibitem [{\citenamefont {Mart\'{\i}nez-Sala}\ \emph
  {et~al.}(1995)\citenamefont {Mart\'{\i}nez-Sala}, \citenamefont {Sancho},
  \citenamefont {S\'{a}nchez}, \citenamefont {G\'{o}mez}, \citenamefont
  {Llinares},\ and\ \citenamefont {Meseguer}}]{martinez-sala_sound_1995}%
  \BibitemOpen
  \bibfield  {author} {\bibinfo {author} {\bibfnamefont {R.}~\bibnamefont
  {Mart\'{\i}nez-Sala}}, \bibinfo {author} {\bibfnamefont {J.}~\bibnamefont
  {Sancho}}, \bibinfo {author} {\bibfnamefont {J.~V.}\ \bibnamefont
  {S\'{a}nchez}}, \bibinfo {author} {\bibfnamefont {V.}~\bibnamefont
  {G\'{o}mez}}, \bibinfo {author} {\bibfnamefont {J.}~\bibnamefont {Llinares}},
  \ and\ \bibinfo {author} {\bibfnamefont {F.}~\bibnamefont {Meseguer}},\
  }\href@noop {} {\bibfield  {journal} {\bibinfo  {journal} {Nature}\ }\textbf
  {\bibinfo {volume} {378}},\ \bibinfo {pages} {241} (\bibinfo {year}
  {1995})}\BibitemShut {NoStop}%
\bibitem [{\citenamefont {Liu}\ \emph {et~al.}(2000)\citenamefont {Liu},
  \citenamefont {Zhang}, \citenamefont {Mao}, \citenamefont {Zhu},
  \citenamefont {Yang}, \citenamefont {Chan},\ and\ \citenamefont
  {Sheng}}]{liu_locally_2000}%
  \BibitemOpen
  \bibfield  {author} {\bibinfo {author} {\bibfnamefont {Z.}~\bibnamefont
  {Liu}}, \bibinfo {author} {\bibfnamefont {X.}~\bibnamefont {Zhang}}, \bibinfo
  {author} {\bibfnamefont {Y.}~\bibnamefont {Mao}}, \bibinfo {author}
  {\bibfnamefont {Y.~Y.}\ \bibnamefont {Zhu}}, \bibinfo {author} {\bibfnamefont
  {Z.}~\bibnamefont {Yang}}, \bibinfo {author} {\bibfnamefont {C.~T.}\
  \bibnamefont {Chan}}, \ and\ \bibinfo {author} {\bibfnamefont
  {P.}~\bibnamefont {Sheng}},\ }\href@noop {} {\bibfield  {journal} {\bibinfo
  {journal} {Science}\ }\textbf {\bibinfo {volume} {289}},\ \bibinfo {pages}
  {1734} (\bibinfo {year} {2000})}\BibitemShut {NoStop}%
\bibitem [{\citenamefont {Khelif}\ \emph {et~al.}(2003)\citenamefont {Khelif},
  \citenamefont {Choujaa}, \citenamefont {Djafari-Rouhani}, \citenamefont
  {Wilm}, \citenamefont {Ballandras},\ and\ \citenamefont
  {Laude}}]{khelif_trapping_2003}%
  \BibitemOpen
  \bibfield  {author} {\bibinfo {author} {\bibfnamefont {A.}~\bibnamefont
  {Khelif}}, \bibinfo {author} {\bibfnamefont {A.}~\bibnamefont {Choujaa}},
  \bibinfo {author} {\bibfnamefont {B.}~\bibnamefont {Djafari-Rouhani}},
  \bibinfo {author} {\bibfnamefont {M.}~\bibnamefont {Wilm}}, \bibinfo {author}
  {\bibfnamefont {S.}~\bibnamefont {Ballandras}}, \ and\ \bibinfo {author}
  {\bibfnamefont {V.}~\bibnamefont {Laude}},\ }\href@noop {} {\bibfield
  {journal} {\bibinfo  {journal} {Physical Review B}\ }\textbf {\bibinfo
  {volume} {68}},\ \bibinfo {pages} {214301} (\bibinfo {year}
  {2003})}\BibitemShut {NoStop}%
\bibitem [{\citenamefont {Pennec}\ \emph {et~al.}(2010)\citenamefont {Pennec},
  \citenamefont {Vasseur}, \citenamefont {Djafari-Rouhani}, \citenamefont
  {Dobrzy\'{n}ski},\ and\ \citenamefont
  {Deymier}}]{pennec_two-dimensional_2010}%
  \BibitemOpen
  \bibfield  {author} {\bibinfo {author} {\bibfnamefont {Y.}~\bibnamefont
  {Pennec}}, \bibinfo {author} {\bibfnamefont {J.~O.}\ \bibnamefont {Vasseur}},
  \bibinfo {author} {\bibfnamefont {B.}~\bibnamefont {Djafari-Rouhani}},
  \bibinfo {author} {\bibfnamefont {L.}~\bibnamefont {Dobrzy\'{n}ski}}, \ and\
  \bibinfo {author} {\bibfnamefont {P.~A.}\ \bibnamefont {Deymier}},\
  }\href@noop {} {\bibfield  {journal} {\bibinfo  {journal} {Surface Science
  Reports}\ }\textbf {\bibinfo {volume} {65}},\ \bibinfo {pages} {229}
  (\bibinfo {year} {2010})}\BibitemShut {NoStop}%
\bibitem [{\citenamefont {Zhu}\ \emph {et~al.}(2011)\citenamefont {Zhu},
  \citenamefont {Christensen}, \citenamefont {Jung}, \citenamefont
  {Martin-Moreno}, \citenamefont {Yin}, \citenamefont {Fok}, \citenamefont
  {Zhang},\ and\ \citenamefont {Garcia-Vidal}}]{zhu_holey-structured_2011}%
  \BibitemOpen
  \bibfield  {author} {\bibinfo {author} {\bibfnamefont {J.}~\bibnamefont
  {Zhu}}, \bibinfo {author} {\bibfnamefont {J.}~\bibnamefont {Christensen}},
  \bibinfo {author} {\bibfnamefont {J.}~\bibnamefont {Jung}}, \bibinfo {author}
  {\bibfnamefont {L.}~\bibnamefont {Martin-Moreno}}, \bibinfo {author}
  {\bibfnamefont {X.}~\bibnamefont {Yin}}, \bibinfo {author} {\bibfnamefont
  {L.}~\bibnamefont {Fok}}, \bibinfo {author} {\bibfnamefont {X.}~\bibnamefont
  {Zhang}}, \ and\ \bibinfo {author} {\bibfnamefont {F.~J.}\ \bibnamefont
  {Garcia-Vidal}},\ }\href@noop {} {\bibfield  {journal} {\bibinfo  {journal}
  {Nature Physics}\ }\textbf {\bibinfo {volume} {7}},\ \bibinfo {pages} {52}
  (\bibinfo {year} {2011})}\BibitemShut {NoStop}%
\bibitem [{\citenamefont {Yang}\ \emph {et~al.}(2014)\citenamefont {Yang},
  \citenamefont {Yang},\ and\ \citenamefont {Li}}]{yang_extreme_2014}%
  \BibitemOpen
  \bibfield  {author} {\bibinfo {author} {\bibfnamefont {L.}~\bibnamefont
  {Yang}}, \bibinfo {author} {\bibfnamefont {N.}~\bibnamefont {Yang}}, \ and\
  \bibinfo {author} {\bibfnamefont {B.}~\bibnamefont {Li}},\ }\href@noop {}
  {\bibfield  {journal} {\bibinfo  {journal} {Nano Letters}\ }\textbf {\bibinfo
  {volume} {14}},\ \bibinfo {pages} {1734} (\bibinfo {year}
  {2014})}\BibitemShut {NoStop}%
\bibitem [{\citenamefont {Xu}\ and\ \citenamefont
  {Yuan}(2018)}]{xu_implementation_2018}%
  \BibitemOpen
  \bibfield  {author} {\bibinfo {author} {\bibfnamefont {Z.}~\bibnamefont
  {Xu}}\ and\ \bibinfo {author} {\bibfnamefont {Y.~J.}\ \bibnamefont {Yuan}},\
  }\href@noop {} {\bibfield  {journal} {\bibinfo  {journal} {Biosensors and
  Bioelectronics}\ }\textbf {\bibinfo {volume} {99}},\ \bibinfo {pages} {500}
  (\bibinfo {year} {2018})}\BibitemShut {NoStop}%
\bibitem [{\citenamefont {Wu}\ \emph {et~al.}(2009)\citenamefont {Wu},
  \citenamefont {Wu},\ and\ \citenamefont {Hsu}}]{wu_waveguiding_2009}%
  \BibitemOpen
  \bibfield  {author} {\bibinfo {author} {\bibfnamefont {T.-C.}\ \bibnamefont
  {Wu}}, \bibinfo {author} {\bibfnamefont {T.-T.}\ \bibnamefont {Wu}}, \ and\
  \bibinfo {author} {\bibfnamefont {J.-C.}\ \bibnamefont {Hsu}},\ }\href@noop
  {} {\bibfield  {journal} {\bibinfo  {journal} {Physical Review B}\ }\textbf
  {\bibinfo {volume} {79}} (\bibinfo {year} {2009})}\BibitemShut {NoStop}%
\bibitem [{\citenamefont {Liang}\ \emph {et~al.}(2010)\citenamefont {Liang},
  \citenamefont {Guo}, \citenamefont {Tu}, \citenamefont {Zhang},\ and\
  \citenamefont {Cheng}}]{liang_acoustic_2010}%
  \BibitemOpen
  \bibfield  {author} {\bibinfo {author} {\bibfnamefont {B.}~\bibnamefont
  {Liang}}, \bibinfo {author} {\bibfnamefont {X.~S.}\ \bibnamefont {Guo}},
  \bibinfo {author} {\bibfnamefont {J.}~\bibnamefont {Tu}}, \bibinfo {author}
  {\bibfnamefont {D.}~\bibnamefont {Zhang}}, \ and\ \bibinfo {author}
  {\bibfnamefont {J.~C.}\ \bibnamefont {Cheng}},\ }\href@noop {} {\bibfield
  {journal} {\bibinfo  {journal} {Nature Materials}\ }\textbf {\bibinfo
  {volume} {9}},\ \bibinfo {pages} {989} (\bibinfo {year} {2010})}\BibitemShut
  {NoStop}%
\bibitem [{\citenamefont {Meseguer}\ \emph {et~al.}(1999)\citenamefont
  {Meseguer}, \citenamefont {Holgado}, \citenamefont {Caballero}, \citenamefont
  {Benaches}, \citenamefont {S\'{a}nchez-Dehesa}, \citenamefont {L\'{o}pez},\
  and\ \citenamefont {Llinares}}]{meseguer_rayleigh-wave_1999}%
  \BibitemOpen
  \bibfield  {author} {\bibinfo {author} {\bibfnamefont {F.}~\bibnamefont
  {Meseguer}}, \bibinfo {author} {\bibfnamefont {M.}~\bibnamefont {Holgado}},
  \bibinfo {author} {\bibfnamefont {D.}~\bibnamefont {Caballero}}, \bibinfo
  {author} {\bibfnamefont {N.}~\bibnamefont {Benaches}}, \bibinfo {author}
  {\bibfnamefont {J.}~\bibnamefont {S\'{a}nchez-Dehesa}}, \bibinfo {author}
  {\bibfnamefont {C.}~\bibnamefont {L\'{o}pez}}, \ and\ \bibinfo {author}
  {\bibfnamefont {J.}~\bibnamefont {Llinares}},\ }\href@noop {} {\bibfield
  {journal} {\bibinfo  {journal} {Physical Review B}\ }\textbf {\bibinfo
  {volume} {59}},\ \bibinfo {pages} {12169} (\bibinfo {year}
  {1999})}\BibitemShut {NoStop}%
\bibitem [{\citenamefont {Olsson~III}\ and\ \citenamefont
  {El-Kady}(2009)}]{olsson_iii_microfabricated_2009}%
  \BibitemOpen
  \bibfield  {author} {\bibinfo {author} {\bibfnamefont {R.~H.}\ \bibnamefont
  {Olsson~III}}\ and\ \bibinfo {author} {\bibfnamefont {I.}~\bibnamefont
  {El-Kady}},\ }\href@noop {} {\bibfield  {journal} {\bibinfo  {journal}
  {Measurement Science and Technology}\ }\textbf {\bibinfo {volume} {20}},\
  \bibinfo {pages} {012002} (\bibinfo {year} {2009})}\BibitemShut {NoStop}%
\bibitem [{\citenamefont {Binci}\ \emph {et~al.}(2016)\citenamefont {Binci},
  \citenamefont {Tu}, \citenamefont {Zhu},\ and\ \citenamefont
  {Lee}}]{binci_planar_2016}%
  \BibitemOpen
  \bibfield  {author} {\bibinfo {author} {\bibfnamefont {L.}~\bibnamefont
  {Binci}}, \bibinfo {author} {\bibfnamefont {C.}~\bibnamefont {Tu}}, \bibinfo
  {author} {\bibfnamefont {H.}~\bibnamefont {Zhu}}, \ and\ \bibinfo {author}
  {\bibfnamefont {J.~E.-Y.}\ \bibnamefont {Lee}},\ }\href@noop {} {\bibfield
  {journal} {\bibinfo  {journal} {Applied Physics Letters}\ }\textbf {\bibinfo
  {volume} {109}},\ \bibinfo {pages} {203501} (\bibinfo {year}
  {2016})}\BibitemShut {NoStop}%
\bibitem [{\citenamefont {Talbi}\ \emph {et~al.}(2006)\citenamefont {Talbi},
  \citenamefont {Sarry}, \citenamefont {Elhakiki}, \citenamefont {Brizoual},
  \citenamefont {Elmazria}, \citenamefont {Nicolay},\ and\ \citenamefont
  {Alnot}}]{talbi_zno/quartz_2006}%
  \BibitemOpen
  \bibfield  {author} {\bibinfo {author} {\bibfnamefont {A.}~\bibnamefont
  {Talbi}}, \bibinfo {author} {\bibfnamefont {F.}~\bibnamefont {Sarry}},
  \bibinfo {author} {\bibfnamefont {M.}~\bibnamefont {Elhakiki}}, \bibinfo
  {author} {\bibfnamefont {L.~L.}\ \bibnamefont {Brizoual}}, \bibinfo {author}
  {\bibfnamefont {O.}~\bibnamefont {Elmazria}}, \bibinfo {author}
  {\bibfnamefont {P.}~\bibnamefont {Nicolay}}, \ and\ \bibinfo {author}
  {\bibfnamefont {P.}~\bibnamefont {Alnot}},\ }\href@noop {} {\bibfield
  {journal} {\bibinfo  {journal} {Sensors and Actuators A: Physical}\ }\textbf
  {\bibinfo {volume} {128}},\ \bibinfo {pages} {78} (\bibinfo {year}
  {2006})}\BibitemShut {NoStop}%
\bibitem [{\citenamefont {Ke}\ \emph {et~al.}(2011)\citenamefont {Ke},
  \citenamefont {Zubtsov},\ and\ \citenamefont
  {Lucklum}}]{ke_sub-wavelength_2011}%
  \BibitemOpen
  \bibfield  {author} {\bibinfo {author} {\bibfnamefont {M.}~\bibnamefont
  {Ke}}, \bibinfo {author} {\bibfnamefont {M.}~\bibnamefont {Zubtsov}}, \ and\
  \bibinfo {author} {\bibfnamefont {R.}~\bibnamefont {Lucklum}},\ }\href@noop
  {} {\bibfield  {journal} {\bibinfo  {journal} {Journal of Applied Physics}\
  }\textbf {\bibinfo {volume} {110}},\ \bibinfo {pages} {026101} (\bibinfo
  {year} {2011})}\BibitemShut {NoStop}%
\bibitem [{\citenamefont {Salman}\ \emph {et~al.}(2015)\citenamefont {Salman},
  \citenamefont {Kaya}, \citenamefont {Cicek},\ and\ \citenamefont
  {Ulug}}]{salman_low-concentration_2015}%
  \BibitemOpen
  \bibfield  {author} {\bibinfo {author} {\bibfnamefont {A.}~\bibnamefont
  {Salman}}, \bibinfo {author} {\bibfnamefont {O.~A.}\ \bibnamefont {Kaya}},
  \bibinfo {author} {\bibfnamefont {A.}~\bibnamefont {Cicek}}, \ and\ \bibinfo
  {author} {\bibfnamefont {B.}~\bibnamefont {Ulug}},\ }\href@noop {} {\bibfield
   {journal} {\bibinfo  {journal} {Journal of Physics D: Applied Physics}\
  }\textbf {\bibinfo {volume} {48}},\ \bibinfo {pages} {255301} (\bibinfo
  {year} {2015})}\BibitemShut {NoStop}%
\bibitem [{\citenamefont {Yu}\ \emph {et~al.}(2010)\citenamefont {Yu},
  \citenamefont {Mitrovic}, \citenamefont {Tham}, \citenamefont {Varghese},\
  and\ \citenamefont {Heath}}]{yu_reduction_2010}%
  \BibitemOpen
  \bibfield  {author} {\bibinfo {author} {\bibfnamefont {J.-K.}\ \bibnamefont
  {Yu}}, \bibinfo {author} {\bibfnamefont {S.}~\bibnamefont {Mitrovic}},
  \bibinfo {author} {\bibfnamefont {D.}~\bibnamefont {Tham}}, \bibinfo {author}
  {\bibfnamefont {J.}~\bibnamefont {Varghese}}, \ and\ \bibinfo {author}
  {\bibfnamefont {J.~R.}\ \bibnamefont {Heath}},\ }\href@noop {} {\bibfield
  {journal} {\bibinfo  {journal} {Nature Nanotechnology}\ }\textbf {\bibinfo
  {volume} {5}},\ \bibinfo {pages} {718} (\bibinfo {year} {2010})}\BibitemShut
  {NoStop}%
\bibitem [{\citenamefont {Zen}\ \emph {et~al.}(2014)\citenamefont {Zen},
  \citenamefont {Puurtinen}, \citenamefont {Isotalo}, \citenamefont
  {Chaudhuri},\ and\ \citenamefont {Maasilta}}]{zen_engineering_2014}%
  \BibitemOpen
  \bibfield  {author} {\bibinfo {author} {\bibfnamefont {N.}~\bibnamefont
  {Zen}}, \bibinfo {author} {\bibfnamefont {T.~A.}\ \bibnamefont {Puurtinen}},
  \bibinfo {author} {\bibfnamefont {T.~J.}\ \bibnamefont {Isotalo}}, \bibinfo
  {author} {\bibfnamefont {S.}~\bibnamefont {Chaudhuri}}, \ and\ \bibinfo
  {author} {\bibfnamefont {I.~J.}\ \bibnamefont {Maasilta}},\ }\href@noop {}
  {\bibfield  {journal} {\bibinfo  {journal} {Nature Communications}\ }\textbf
  {\bibinfo {volume} {5}} (\bibinfo {year} {2014})}\BibitemShut {NoStop}%
\bibitem [{\citenamefont {Zhang}\ and\ \citenamefont
  {Liu}(2004)}]{zhang_negative_2004}%
  \BibitemOpen
  \bibfield  {author} {\bibinfo {author} {\bibfnamefont {X.}~\bibnamefont
  {Zhang}}\ and\ \bibinfo {author} {\bibfnamefont {Z.}~\bibnamefont {Liu}},\
  }\href@noop {} {\bibfield  {journal} {\bibinfo  {journal} {Applied Physics
  Letters}\ }\textbf {\bibinfo {volume} {85}},\ \bibinfo {pages} {341}
  (\bibinfo {year} {2004})}\BibitemShut {NoStop}%
\bibitem [{\citenamefont {Profunser}\ \emph {et~al.}(2009)\citenamefont
  {Profunser}, \citenamefont {Muramoto}, \citenamefont {Matsuda}, \citenamefont
  {Wright},\ and\ \citenamefont {Lang}}]{profunser_dynamic_2009}%
  \BibitemOpen
  \bibfield  {author} {\bibinfo {author} {\bibfnamefont {D.~M.}\ \bibnamefont
  {Profunser}}, \bibinfo {author} {\bibfnamefont {E.}~\bibnamefont {Muramoto}},
  \bibinfo {author} {\bibfnamefont {O.}~\bibnamefont {Matsuda}}, \bibinfo
  {author} {\bibfnamefont {O.~B.}\ \bibnamefont {Wright}}, \ and\ \bibinfo
  {author} {\bibfnamefont {U.}~\bibnamefont {Lang}},\ }\href@noop {} {\bibfield
   {journal} {\bibinfo  {journal} {Physical Review B}\ }\textbf {\bibinfo
  {volume} {80}},\ \bibinfo {pages} {014301} (\bibinfo {year}
  {2009})}\BibitemShut {NoStop}%
\bibitem [{\citenamefont {Narayana}\ and\ \citenamefont
  {Sato}(2012)}]{narayana_heat_2012}%
  \BibitemOpen
  \bibfield  {author} {\bibinfo {author} {\bibfnamefont {S.}~\bibnamefont
  {Narayana}}\ and\ \bibinfo {author} {\bibfnamefont {Y.}~\bibnamefont
  {Sato}},\ }\href@noop {} {\bibfield  {journal} {\bibinfo  {journal} {Physical
  Review Letters}\ }\textbf {\bibinfo {volume} {108}},\ \bibinfo {pages}
  {214303} (\bibinfo {year} {2012})}\BibitemShut {NoStop}%
\bibitem [{\citenamefont {Li}\ \emph {et~al.}(2015)\citenamefont {Li},
  \citenamefont {Wu}, \citenamefont {Zhong}, \citenamefont {Yao},\ and\
  \citenamefont {Zhang}}]{li_acoustic_2015}%
  \BibitemOpen
  \bibfield  {author} {\bibinfo {author} {\bibfnamefont {J.}~\bibnamefont
  {Li}}, \bibinfo {author} {\bibfnamefont {F.}~\bibnamefont {Wu}}, \bibinfo
  {author} {\bibfnamefont {H.}~\bibnamefont {Zhong}}, \bibinfo {author}
  {\bibfnamefont {Y.}~\bibnamefont {Yao}}, \ and\ \bibinfo {author}
  {\bibfnamefont {X.}~\bibnamefont {Zhang}},\ }\href@noop {} {\bibfield
  {journal} {\bibinfo  {journal} {Journal of Applied Physics}\ }\textbf
  {\bibinfo {volume} {118}},\ \bibinfo {pages} {144903} (\bibinfo {year}
  {2015})}\BibitemShut {NoStop}%
\bibitem [{\citenamefont {Guo}\ \emph {et~al.}(2017)\citenamefont {Guo},
  \citenamefont {Brick}, \citenamefont {Gro{\ss}mann}, \citenamefont
  {Hettich},\ and\ \citenamefont {Dekorsy}}]{guo_acoustic_2017}%
  \BibitemOpen
  \bibfield  {author} {\bibinfo {author} {\bibfnamefont {Y.}~\bibnamefont
  {Guo}}, \bibinfo {author} {\bibfnamefont {D.}~\bibnamefont {Brick}}, \bibinfo
  {author} {\bibfnamefont {M.}~\bibnamefont {Gro{\ss}mann}}, \bibinfo {author}
  {\bibfnamefont {M.}~\bibnamefont {Hettich}}, \ and\ \bibinfo {author}
  {\bibfnamefont {T.}~\bibnamefont {Dekorsy}},\ }\href@noop {} {\bibfield
  {journal} {\bibinfo  {journal} {Applied Physics Letters}\ }\textbf {\bibinfo
  {volume} {110}},\ \bibinfo {pages} {031904} (\bibinfo {year}
  {2017})}\BibitemShut {NoStop}%
\bibitem [{\citenamefont {Li}\ \emph {et~al.}(2011)\citenamefont {Li},
  \citenamefont {Zhao}, \citenamefont {Alu},\ and\ \citenamefont
  {Yu}}]{li_experimetal_2011}%
  \BibitemOpen
  \bibfield  {author} {\bibinfo {author} {\bibfnamefont {P.-C.}\ \bibnamefont
  {Li}}, \bibinfo {author} {\bibfnamefont {Y.}~\bibnamefont {Zhao}}, \bibinfo
  {author} {\bibfnamefont {A.}~\bibnamefont {Alu}}, \ and\ \bibinfo {author}
  {\bibfnamefont {E.~T.}\ \bibnamefont {Yu}},\ }\href@noop {} {\bibfield
  {journal} {\bibinfo  {journal} {Applied Physics Letters}\ }\textbf {\bibinfo
  {volume} {99}},\ \bibinfo {pages} {221106} (\bibinfo {year}
  {2011})}\BibitemShut {NoStop}%
\bibitem [{\citenamefont {Yu}\ and\ \citenamefont
  {Capasso}(2014)}]{yu_flat_2014}%
  \BibitemOpen
  \bibfield  {author} {\bibinfo {author} {\bibfnamefont {N.}~\bibnamefont
  {Yu}}\ and\ \bibinfo {author} {\bibfnamefont {F.}~\bibnamefont {Capasso}},\
  }\href@noop {} {\bibfield  {journal} {\bibinfo  {journal} {Nature Materials}\
  }\textbf {\bibinfo {volume} {13}},\ \bibinfo {pages} {139} (\bibinfo {year}
  {2014})}\BibitemShut {NoStop}%
\bibitem [{\citenamefont {Oudich}\ \emph {et~al.}(2018)\citenamefont {Oudich},
  \citenamefont {Djafari-Rouhani}, \citenamefont {Bonello}, \citenamefont
  {Pennec}, \citenamefont {Hemaidia}, \citenamefont {Sarry},\ and\
  \citenamefont {Beyssen}}]{oudich_rayleigh_2018}%
  \BibitemOpen
  \bibfield  {author} {\bibinfo {author} {\bibfnamefont {M.}~\bibnamefont
  {Oudich}}, \bibinfo {author} {\bibfnamefont {B.}~\bibnamefont
  {Djafari-Rouhani}}, \bibinfo {author} {\bibfnamefont {B.}~\bibnamefont
  {Bonello}}, \bibinfo {author} {\bibfnamefont {Y.}~\bibnamefont {Pennec}},
  \bibinfo {author} {\bibfnamefont {S.}~\bibnamefont {Hemaidia}}, \bibinfo
  {author} {\bibfnamefont {F.}~\bibnamefont {Sarry}}, \ and\ \bibinfo {author}
  {\bibfnamefont {D.}~\bibnamefont {Beyssen}},\ }\href@noop {} {\bibfield
  {journal} {\bibinfo  {journal} {Physical Review Applied}\ }\textbf {\bibinfo
  {volume} {9}},\ \bibinfo {pages} {034013} (\bibinfo {year}
  {2018})}\BibitemShut {NoStop}%
\bibitem [{\citenamefont {Jin}\ \emph {et~al.}(2017{\natexlab{a}})\citenamefont
  {Jin}, \citenamefont {El~Boudouti}, \citenamefont {Pennec},\ and\
  \citenamefont {Djafari-Rouhani}}]{jin_tunable_2017}%
  \BibitemOpen
  \bibfield  {author} {\bibinfo {author} {\bibfnamefont {Y.}~\bibnamefont
  {Jin}}, \bibinfo {author} {\bibfnamefont {E.~H.}\ \bibnamefont
  {El~Boudouti}}, \bibinfo {author} {\bibfnamefont {Y.}~\bibnamefont {Pennec}},
  \ and\ \bibinfo {author} {\bibfnamefont {B.}~\bibnamefont
  {Djafari-Rouhani}},\ }\href@noop {} {\bibfield  {journal} {\bibinfo
  {journal} {Journal of Physics D: Applied Physics}\ }\textbf {\bibinfo
  {volume} {50}},\ \bibinfo {pages} {425304} (\bibinfo {year}
  {2017}{\natexlab{a}})}\BibitemShut {NoStop}%
\bibitem [{\citenamefont {Jin}\ \emph {et~al.}(2017{\natexlab{b}})\citenamefont
  {Jin}, \citenamefont {Bonello}, \citenamefont {Moiseyenko}, \citenamefont
  {Pennec}, \citenamefont {Boyko},\ and\ \citenamefont
  {Djafari-Rouhani}}]{jin_acoustic_2017}%
  \BibitemOpen
  \bibfield  {author} {\bibinfo {author} {\bibfnamefont {Y.}~\bibnamefont
  {Jin}}, \bibinfo {author} {\bibfnamefont {B.}~\bibnamefont {Bonello}},
  \bibinfo {author} {\bibfnamefont {R.~P.}\ \bibnamefont {Moiseyenko}},
  \bibinfo {author} {\bibfnamefont {Y.}~\bibnamefont {Pennec}}, \bibinfo
  {author} {\bibfnamefont {O.}~\bibnamefont {Boyko}}, \ and\ \bibinfo {author}
  {\bibfnamefont {B.}~\bibnamefont {Djafari-Rouhani}},\ }\href@noop {}
  {\bibfield  {journal} {\bibinfo  {journal} {Physical Review B}\ }\textbf
  {\bibinfo {volume} {96}},\ \bibinfo {pages} {104311} (\bibinfo {year}
  {2017}{\natexlab{b}})}\BibitemShut {NoStop}%
\bibitem [{\citenamefont {Liu}\ \emph {et~al.}(2019)\citenamefont {Liu},
  \citenamefont {Talbi}, \citenamefont {Djafari-Rouhani}, \citenamefont
  {El~Boudouti}, \citenamefont {Dobohlavova}, \citenamefont {Mortet},
  \citenamefont {Bou~Matar},\ and\ \citenamefont
  {Pernod}}]{liu_interaction_2019}%
  \BibitemOpen
  \bibfield  {author} {\bibinfo {author} {\bibfnamefont {Y.}~\bibnamefont
  {Liu}}, \bibinfo {author} {\bibfnamefont {A.}~\bibnamefont {Talbi}}, \bibinfo
  {author} {\bibfnamefont {B.}~\bibnamefont {Djafari-Rouhani}}, \bibinfo
  {author} {\bibfnamefont {E.~H.}\ \bibnamefont {El~Boudouti}}, \bibinfo
  {author} {\bibfnamefont {L.}~\bibnamefont {Dobohlavova}}, \bibinfo {author}
  {\bibfnamefont {V.}~\bibnamefont {Mortet}}, \bibinfo {author} {\bibfnamefont
  {O.}~\bibnamefont {Bou~Matar}}, \ and\ \bibinfo {author} {\bibfnamefont
  {P.}~\bibnamefont {Pernod}},\ }\href@noop {} {\bibfield  {journal} {\bibinfo
  {journal} {Physics Letters A}\ }\textbf {\bibinfo {volume} {383}},\ \bibinfo
  {pages} {1502} (\bibinfo {year} {2019})}\BibitemShut {NoStop}%
\bibitem [{\citenamefont {Lucklum}\ and\ \citenamefont
  {Li}(2009)}]{lucklum_phononic_2009}%
  \BibitemOpen
  \bibfield  {author} {\bibinfo {author} {\bibfnamefont {R.}~\bibnamefont
  {Lucklum}}\ and\ \bibinfo {author} {\bibfnamefont {J.}~\bibnamefont {Li}},\
  }\href@noop {} {\bibfield  {journal} {\bibinfo  {journal} {Measurement
  Science and Technology}\ }\textbf {\bibinfo {volume} {20}},\ \bibinfo {pages}
  {124014} (\bibinfo {year} {2009})}\BibitemShut {NoStop}%
\bibitem [{\citenamefont {Liu}\ \emph {et~al.}(2014{\natexlab{a}})\citenamefont
  {Liu}, \citenamefont {Lin}, \citenamefont {Tsai}, \citenamefont {Ono},
  \citenamefont {Tanaka},\ and\ \citenamefont {Wu}}]{liu_evidence_2014}%
  \BibitemOpen
  \bibfield  {author} {\bibinfo {author} {\bibfnamefont {T.-W.}\ \bibnamefont
  {Liu}}, \bibinfo {author} {\bibfnamefont {Y.-C.}\ \bibnamefont {Lin}},
  \bibinfo {author} {\bibfnamefont {Y.-C.}\ \bibnamefont {Tsai}}, \bibinfo
  {author} {\bibfnamefont {T.}~\bibnamefont {Ono}}, \bibinfo {author}
  {\bibfnamefont {S.}~\bibnamefont {Tanaka}}, \ and\ \bibinfo {author}
  {\bibfnamefont {T.-T.}\ \bibnamefont {Wu}},\ }\href@noop {} {\bibfield
  {journal} {\bibinfo  {journal} {Applied Physics Letters}\ }\textbf {\bibinfo
  {volume} {104}},\ \bibinfo {pages} {181905} (\bibinfo {year}
  {2014}{\natexlab{a}})}\BibitemShut {NoStop}%
\bibitem [{\citenamefont {Liu}\ \emph {et~al.}(2014{\natexlab{b}})\citenamefont
  {Liu}, \citenamefont {Tsai}, \citenamefont {Lin}, \citenamefont {Ono},
  \citenamefont {Tanaka},\ and\ \citenamefont {Wu}}]{liu_design_2014}%
  \BibitemOpen
  \bibfield  {author} {\bibinfo {author} {\bibfnamefont {T.-W.}\ \bibnamefont
  {Liu}}, \bibinfo {author} {\bibfnamefont {Y.-C.}\ \bibnamefont {Tsai}},
  \bibinfo {author} {\bibfnamefont {Y.-C.}\ \bibnamefont {Lin}}, \bibinfo
  {author} {\bibfnamefont {T.}~\bibnamefont {Ono}}, \bibinfo {author}
  {\bibfnamefont {S.}~\bibnamefont {Tanaka}}, \ and\ \bibinfo {author}
  {\bibfnamefont {T.-T.}\ \bibnamefont {Wu}},\ }\href@noop {} {\bibfield
  {journal} {\bibinfo  {journal} {AIP Advances}\ }\textbf {\bibinfo {volume}
  {4}},\ \bibinfo {pages} {124201} (\bibinfo {year}
  {2014}{\natexlab{b}})}\BibitemShut {NoStop}%
\bibitem [{\citenamefont {Liu}\ \emph {et~al.}(2018)\citenamefont {Liu},
  \citenamefont {Talbi}, \citenamefont {Pernod},\ and\ \citenamefont
  {Bou~Matar}}]{liu_highly_2018}%
  \BibitemOpen
  \bibfield  {author} {\bibinfo {author} {\bibfnamefont {Y.}~\bibnamefont
  {Liu}}, \bibinfo {author} {\bibfnamefont {A.}~\bibnamefont {Talbi}}, \bibinfo
  {author} {\bibfnamefont {P.}~\bibnamefont {Pernod}}, \ and\ \bibinfo {author}
  {\bibfnamefont {O.}~\bibnamefont {Bou~Matar}},\ }\href@noop {} {\bibfield
  {journal} {\bibinfo  {journal} {Journal of Applied Physics}\ }\textbf
  {\bibinfo {volume} {124}},\ \bibinfo {pages} {145102} (\bibinfo {year}
  {2018})}\BibitemShut {NoStop}%
\bibitem [{\citenamefont {Yankin}\ \emph {et~al.}(2014)\citenamefont {Yankin},
  \citenamefont {Talbi}, \citenamefont {Du}, \citenamefont {Gerbedoen},
  \citenamefont {Preobrazhensky}, \citenamefont {Pernod},\ and\ \citenamefont
  {Bou~Matar}}]{yankin_finite_2014}%
  \BibitemOpen
  \bibfield  {author} {\bibinfo {author} {\bibfnamefont {S.}~\bibnamefont
  {Yankin}}, \bibinfo {author} {\bibfnamefont {A.}~\bibnamefont {Talbi}},
  \bibinfo {author} {\bibfnamefont {Y.}~\bibnamefont {Du}}, \bibinfo {author}
  {\bibfnamefont {J.-C.}\ \bibnamefont {Gerbedoen}}, \bibinfo {author}
  {\bibfnamefont {V.}~\bibnamefont {Preobrazhensky}}, \bibinfo {author}
  {\bibfnamefont {P.}~\bibnamefont {Pernod}}, \ and\ \bibinfo {author}
  {\bibfnamefont {O.}~\bibnamefont {Bou~Matar}},\ }\href@noop {} {\bibfield
  {journal} {\bibinfo  {journal} {Journal of Applied Physics}\ }\textbf
  {\bibinfo {volume} {115}},\ \bibinfo {pages} {244508} (\bibinfo {year}
  {2014})}\BibitemShut {NoStop}%
\bibitem [{\citenamefont {Burnham}\ and\ \citenamefont
  {Anderson}(2002)}]{burnham_model_2002}%
  \BibitemOpen
  \bibinfo {editor} {\bibfnamefont {K.~P.}\ \bibnamefont {Burnham}}\ and\
  \bibinfo {editor} {\bibfnamefont {D.~R.}\ \bibnamefont {Anderson}},\ eds.,\
  \href@noop {} {\emph {\bibinfo {title} {Model Selection and Multimodel
  Inference}}},\ \bibinfo {edition} {2nd}\ ed.\ (\bibinfo  {publisher}
  {Springer-Verlag, New York},\ \bibinfo {year} {2002})\BibitemShut {NoStop}%
\bibitem [{\citenamefont {Talbi}\ \emph {et~al.}(2015)\citenamefont {Talbi},
  \citenamefont {Soltani}, \citenamefont {Rumeau}, \citenamefont {Taylor},
  \citenamefont {Drbohlavova}, \citenamefont {Klimsa}, \citenamefont {Kopecek},
  \citenamefont {Fekete}, \citenamefont {Krecmarova},\ and\ \citenamefont
  {Mortet}}]{talbi_simulation_2015}%
  \BibitemOpen
  \bibfield  {author} {\bibinfo {author} {\bibfnamefont {A.}~\bibnamefont
  {Talbi}}, \bibinfo {author} {\bibfnamefont {A.}~\bibnamefont {Soltani}},
  \bibinfo {author} {\bibfnamefont {A.}~\bibnamefont {Rumeau}}, \bibinfo
  {author} {\bibfnamefont {A.}~\bibnamefont {Taylor}}, \bibinfo {author}
  {\bibfnamefont {L.}~\bibnamefont {Drbohlavova}}, \bibinfo {author}
  {\bibfnamefont {L.}~\bibnamefont {Klimsa}}, \bibinfo {author} {\bibfnamefont
  {J.}~\bibnamefont {Kopecek}}, \bibinfo {author} {\bibfnamefont
  {L.}~\bibnamefont {Fekete}}, \bibinfo {author} {\bibfnamefont
  {M.}~\bibnamefont {Krecmarova}}, \ and\ \bibinfo {author} {\bibfnamefont
  {V.}~\bibnamefont {Mortet}},\ }\href@noop {} {\bibfield  {journal} {\bibinfo
  {journal} {Physica Status Solidi A-Applications and Materials Science}\
  }\textbf {\bibinfo {volume} {212}},\ \bibinfo {pages} {2606} (\bibinfo {year}
  {2015})}\BibitemShut {NoStop}%
\end{thebibliography}%

\end{document}